\documentclass[10pt,letterpaper]{article}
\usepackage[top=0.85in,left=2.75in,footskip=0.75in]{geometry}

\usepackage{labelfig}

\usepackage{amsmath,amssymb}

\usepackage{changepage}

\usepackage[utf8x]{inputenc}

\usepackage{textcomp,marvosym}

\usepackage{cite}

\usepackage{nameref,hyperref}

\usepackage{microtype}
\DisableLigatures[f]{encoding = *, family = * }

\usepackage[table]{xcolor}

\usepackage{array}

\newcolumntype{+}{!{\vrule width 2pt}}

\newlength\savedwidth

\newcommand\thickhline{\noalign{\global\savedwidth\arrayrulewidth\global\arrayrulewidth 2pt}%
\hline
\noalign{\global\arrayrulewidth\savedwidth}}

\raggedright
\usepackage[aboveskip=1pt,labelfont=bf,labelsep=period,justification=raggedright,singlelinecheck=off]{caption}

\makeatletter
\renewcommand{\@biblabel}[1]{\quad#1.}
\makeatother

\date{}

\usepackage{lastpage,fancyhdr,graphicx}
\usepackage{epstopdf}
\fancyheadoffset[L]{2.25in}
\fancyfootoffset[L]{2.25in}
\begin{document}
\vspace*{0.2in}

\begin{flushleft}
{\Large
\textbf\newline{Experimental and Theoretical Study of Magnetohydrodynamic Ship Models} 
}
\newline
\\
David C\'ebron\textsuperscript{1*},
Sylvain Viroulet\textsuperscript{2},
J\'er\'emie Vidal\textsuperscript{1},
Jean-Paul Masson\textsuperscript{1},
Philippe Viroulet\textsuperscript{3}
\\
\bigskip
\textbf{1} Univ. Grenoble Alpes, CNRS, ISTerre, F-38000 Grenoble, France
\\
\textbf{2} School of Mathematics and Manchester Centre for Nonlinear Dynamics, University of Manchester, M13 9PL, Manchester, UK
\\
\textbf{3} Independent Researcher, Quimper, France
\bigskip

%
%





* david.cebron@univ-grenoble-alpes.fr

\end{flushleft}
\section*{Abstract}
Magnetohydrodynamic (MHD) ships represent a clear demonstration of the Lorentz force in fluids, which explains the number of students practicals or exercises described on the web. However, the related literature is rather specific and no complete comparison between theory and typical small scale experiments is currently available. This work provides, in a self-consistent framework, a detailed presentation of the relevant theoretical equations for small MHD ships and experimental measurements for future benchmarks. Theoretical results of the literature are adapted to these simple battery/magnets powered ships moving on salt water. Comparison between theory and experiments are performed to validate each theoretical step such as the Tafel and the Kohlrausch laws, or the predicted ship speed. A successful agreement is obtained without any adjustable parameter. Finally, based on these results, an optimal design is then deduced from the theory. Therefore this work provides a solid theoretical and experimental ground for small scale MHD ships, by presenting in detail several approximations and how they affect the boat efficiency. Moreover, the theory is general enough to be adapted to other contexts, such as large scale ships or industrial flow measurement techniques.


\section*{Introduction} 

Magnetohydrodynamic forces, such as the Lorentz force in fluids, may lead to large scale observations such as the Earth global magnetic field. This field, which orientates any compass, is indeed generated by the strong Lorentz forces present in the Earth liquid core. However, examples of fluid Lorentz forces in our daily life are not common, which explains the difficulties sometimes encountered by students in magnetohydrodynamics to grasp these concepts. In this paper, the action of fluid Lorentz forces is demonstrated by considering a simple setup: the propulsion of a magnetohydrodynamic ship in an electrically conducting fluid. As detailed in the review of \cite{weier2007flow}, this propulsion method, first proposed by \cite{rice1961propulsion,friauf1961electromagnetic,phillips1962prospects}, is attractive in many aspects since this kind of magnetohydrodynamic (or MHD) propulsion does not require any moving parts. MHD propulsion has thus been proposed in seawater for high speed cargo submarines \cite{Way1968}, silent propulsion of naval submarines \cite{swallom1991magnetohydrodynamic,Lin1995}, or high speed ship without any cavitation \cite{nishigaki2000elementary}.

The MHD force propelling the ship can be generated in various ways. The simplest one, the so-called conductive system, imposes both steady magnetic and electric fields. As proposed by \cite{phillips1962prospects} and later by \cite{Khonichev1978}, one could also only impose an unsteady magnetic field, either by using unsteady currents or moving magnets \cite{saji1988fundamental}. These so-called inductive systems are convenient because they do not use electrodes and avoid fluid electrolysis, making the ship more quiet and its maintenance easier. Inductive thrusters have thus been extensively studied, mainly theoretically \cite{Khonichev1980a,Shatrov1981,yakovlev1980theory}. Both inductive and conductive systems can be used either within a duct (internal system), or in the surrounding fluid (external system), thus forming in total four different types of MHD thrusters  \cite{Convert1995}. To test the MHD propulsion, large-scale ships have been built, such as a $3$ m long external conductive submarine reaching a maximum velocity of $0.4\, \textrm{m}.\textrm{s}^{-1}$  in the 60's \cite{Way1968}. Two external conductive ships navigating at a maximum velocity of $0.6\, \textrm{m}.\textrm{s}^{-1}$ were built in the 70's \cite{saji1978basic,iwata1980construction}. Two internal conductive ships were built in the 90's, a $30$ m long ship reached a maximum velocity of $3.4\, \textrm{m}.\textrm{s}^{-1}$  \cite{motora1994development,sasakawa1995superconducting}, and a $3.5$ m long one a maximum speed of $0.68\, \textrm{m}.\textrm{s}^{-1}$ \cite{Yan2000,yan2000development}. These practical tests, complemented by theoretical and numerical studies (e.g. \cite{Khonichev1981,shatrov1985hydrodynamic,pohjavirta1991feasibility,convert1995external} for external conductive systems), have shown that the MHD thruster efficiency remains far below the efficiency of common propulsion devices, mainly because of the weakness of the magnetic fields which can be achieved in practice currently \cite{weier2007flow}. The worldwide research interest in MHD propulsion has thus decreased, but the simplicity and the fascination for this kind of propulsion has recently been used as an educational tool \cite{Font2004}. Moreover, even if a submarine propelled by a MHD thruster is still nowadays only applicable in Hollywood movie inspired from Tom Clancy's novel \cite{Clancy1984}, flow control or measurements using magnetohydrodynamic forces is becoming more and more important with the improvement of technologies. Indeed, a contact-less electromagnetic flow measurement technique called Lorentz force flowmeter can be used to measure flow velocities in hot or highly corrosive liquids such as liquid metals or acids \cite{Baumgartl1993,Thess2006,Priede2011}.

This work aims at studying a self-propelled MHD ship model based on an internal conductive square thruster. Such a study imposes to consider a whole range of different aspects of Physics (e.g. fluid mechanics, electrical circuits, electrolysis, etc.) in order to compare experimental measurements  with theoretical predictions. Experiments are performed with a small-scale self-propelled MHD ship model (section \ref{sec:expe}). Then theoretical results available in the literature \cite{Brown1990,Gilbert1991a} are detailed and adapted to MHD ships to investigated the thruster electrical (section \ref{sec:elec}) and magnetic (section \ref{sec:mag}) properties respectively. These preliminary results allow to tackle the study of the MHD ship in section \ref{sec:thruster}. Having obtained the equations governing the ship velocity in section \ref{sec:thruster2}, the complete system of equations (summarized in section \ref{sec:finEQ}) can be analytically solved in particular cases (section \ref{sec:analyTH}) and compared with our measured velocities (section \ref{sec:expeSpeed}). We finally describe how a ship can be optimized for speed using the theory presented in this article, allowing our model to reach a maximum velocity of $0.3\, \textrm{m}.\textrm{s}^{-1}$ (section \ref{sec:DESIGN}). Note that there are very few well-controlled experiments of this kind in the literature. The experimental results presented here provides thus a good comparison point for the theory associated with MHD conductive internal square thrusters.

\section{Experimental setup and method} \label{sec:expe}

Experiments have been performed in a $2\, \textrm{m}$ long, $0.6\, \textrm{m}$ wide and $0.4\, \textrm{m}$ high tank. The water depth remained constant at $h=0.12\, \textrm{m}$ throughout the entire experimental campaign. Table salt was used to change the concentration of NaCl ions in water.
High concentrations of salt are considered in this paper, thus requiring important quantities. For instance, about $42\, \textrm{kg}$ of salt was dissolved into the water tank to reach the concentration $291\, \textrm{kg.m}^{-3}$. Note that the complete dissolution of NaCl requires more and more time as we approach the solubility of NaCl, around $360\, \textrm{kg.m}^{-3}$ in water at $25^{\circ}\, \textrm{C}$.

\begin{figure}[h!]
\centering
 \begin{tabular}{ccc}
\includegraphics[width=8cm]{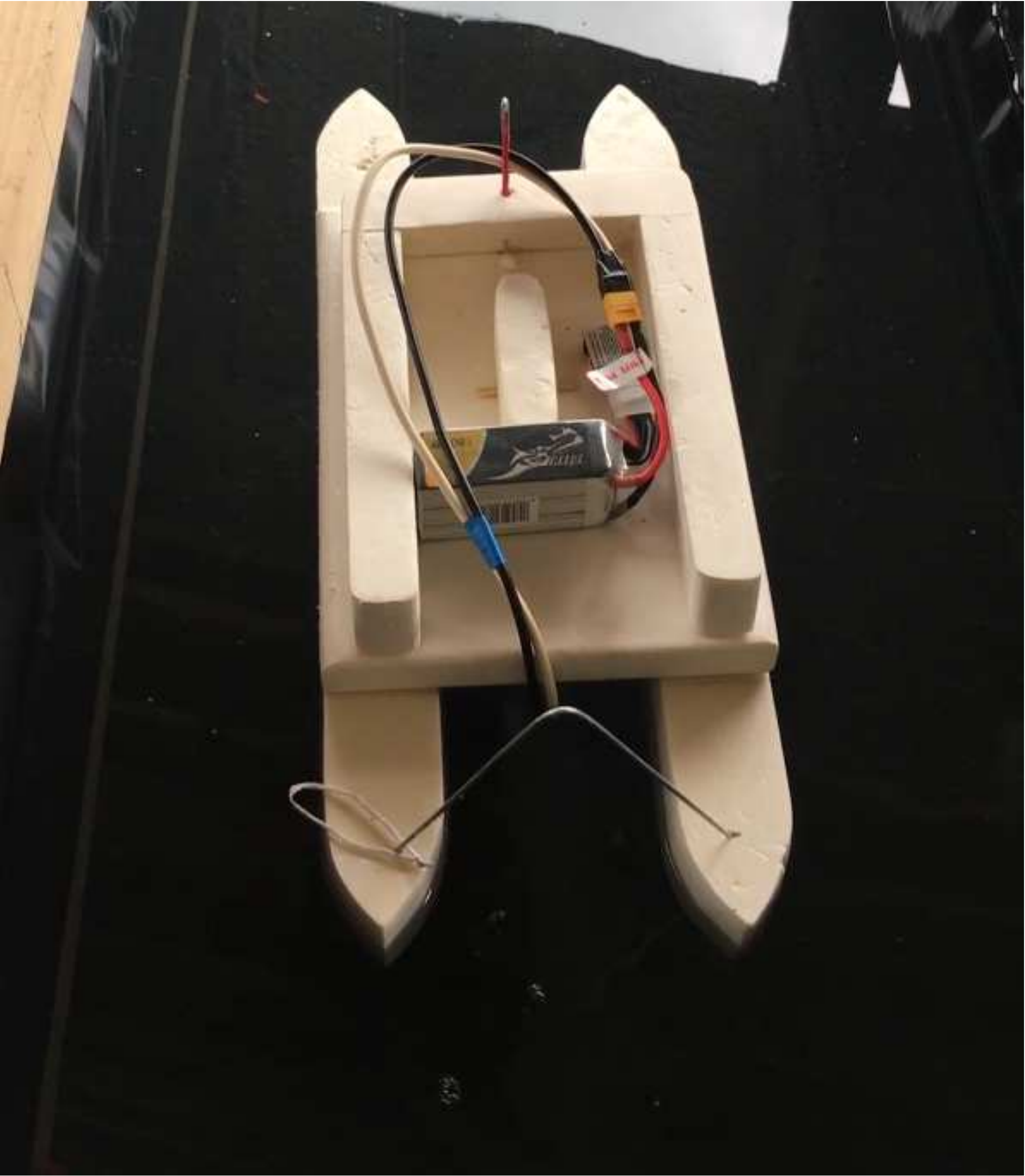} &
 \end{tabular}
  \caption{{\bf Picture of the ship in the experimental tank}. The LiPo 6s battery is placed on top of the boat and connected to the electrode using an XT60 plug.}
  \label{fig01}
\end{figure}

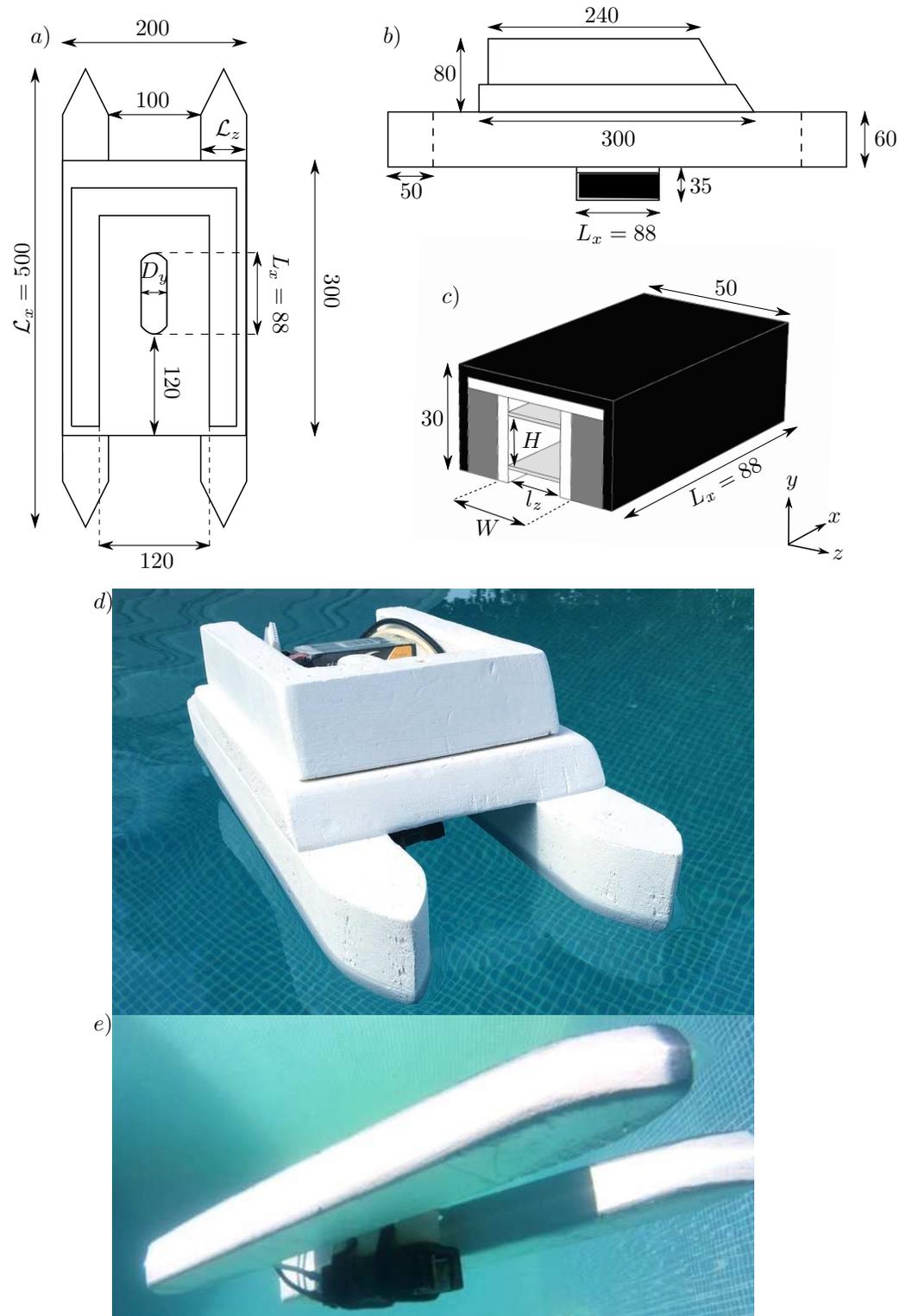
\begin{figure*}
\centering
   \SetLabels
   \L\B (0*0.98) $a)$ \\
   \L\B (-0.02*0.76) {\rotatebox{90}{$\mathcal{L}_x=500$}} \\ 
   \L\B (0.125*0.99) $200$ \\
   \L\B (0.125*0.935) $100$ \\
   \L\B (0.35*0.78) {\rotatebox{-90}{$300$}} \\
   \L\B (0.285*0.76) {\rotatebox{-90}{$L_x=88$}} \\ 
   \L\B (0.16*0.71) {\rotatebox{-90}{$120$}} \\
   \L\B (0.13*0.58) $120$ \\
   \L\B (0.13*0.798) $D_y$ \\  
   \L\B (0.22*0.91) $\mathcal{L}_z$ \\ 
   \L\B (0.42*0.98) $b)$ \\
   \L\B (0.66*1) $240$ \\
   \L\B (0.48*0.955) $80$ \\
   \L\B (0.68*0.905) $300$ \\
   \L\B (0.44*0.865) $50$ \\
   \L\B (0.65*0.83) $L_x=88$ \\ 
   \L\B (0.785*0.8685) $35$ \\
   \L\B (1.005*0.905) $60$ \\
   \L\B (0.49*0.78) $c)$ \\
   \L\B (0.815*0.79) $50$ \\
   \L\B (0.465*0.69) $30$ \\
   \L\B (0.585*0.6725) $H$ \\ 
   \L\B (0.78*0.625) {\rotatebox{28}{$L_x=88$}} \\ 
   \L\B (0.59*0.625) $l_z$ \\      
   \L\B (0.53*0.605) $W$ \\   
   \L\B (0.95*0.615) $x$ \\ 
   \L\B (0.9*0.64) $y$ \\      
   \L\B (0.955*0.587) $z$ \\ 
   \L\B (0.075*0.545) $d)$ \\
   \L\B (0.075*0.22) $e)$ \\
   \endSetLabels

  \strut \AffixLabels{\includegraphics[width=0.97\textwidth]{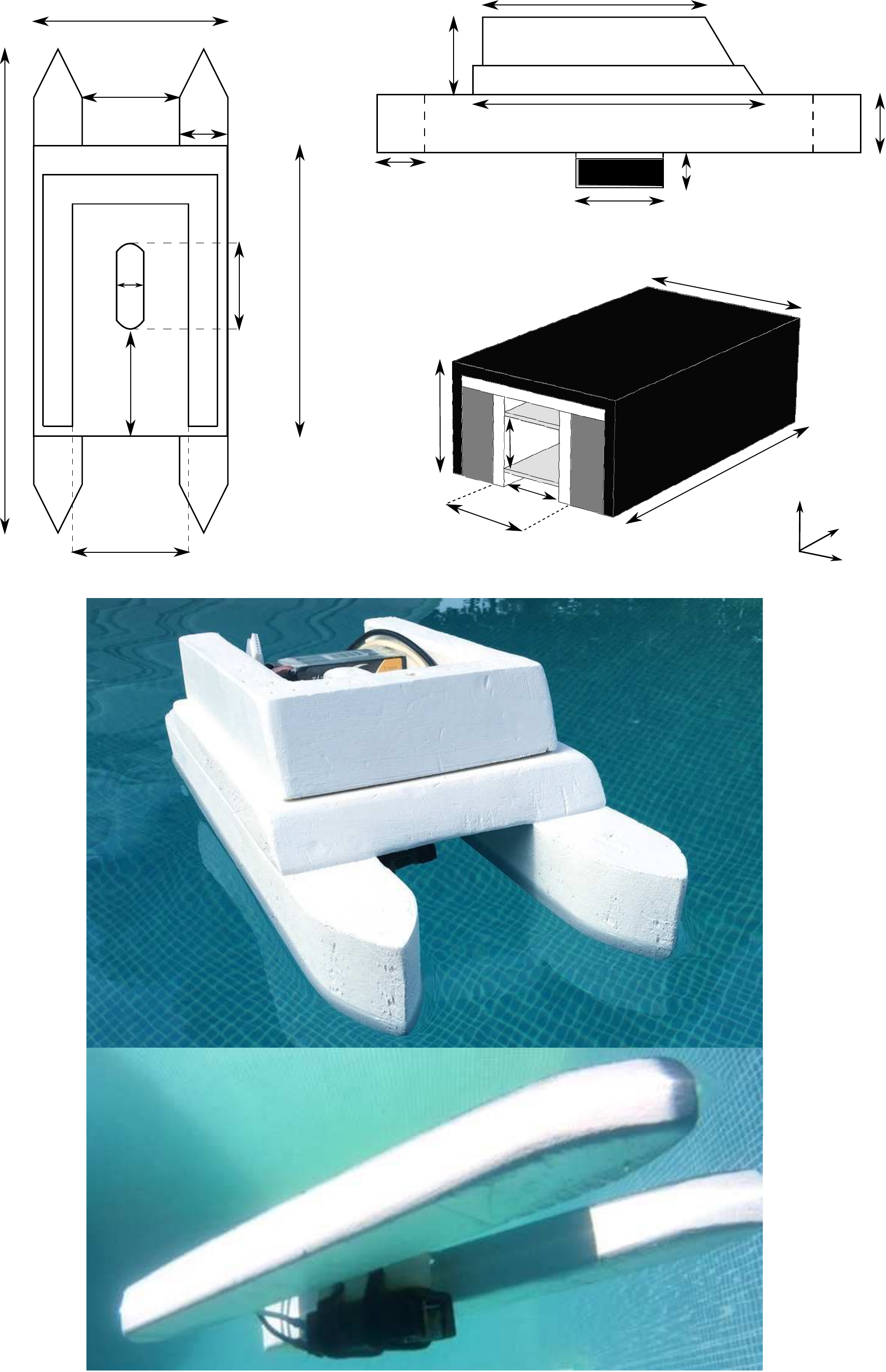}}
  \caption{{\bf Schematic representation of top (a) and side (b) view of the ship (length are in millimeters)}. c) Diagram of the thruster used to propel the ship. The black part represents the magnetic bridge, dark gray part the magnets, light gray part the electrodes and the white part the plastic used to isolate the electric circuit. Pictures (d) and (e) are aerial and underwater view of the ship respectively. The 6s LiPo Battery is visible on picture (d). Note that for aesthetics reasons, pictures have been taken in a swimming pool and not in the actual experimental tank.}
         \label{fig02}
\end{figure*}

The ship is a so-called multihull ship which is at the same time more efficient and more stable than a monohull (Fig~\ref{fig01}). It is made of polystyrene. Each of the floats are $50\, \textrm{cm}$ long, $5\, \textrm{cm}$ wide and $6\, \textrm{cm}$ high, separated by $10\, \textrm{cm}$ between each other (Fig~\ref{fig02}A and B). The aerial part of the ship consists of a $30\, \textrm{cm}$ long, $20\, \textrm{cm}$ large and $3\, \textrm{cm}$ thick platform where $5\, \textrm{cm}$ high edges have been installed to avoid the Lithium-Polymere (LiPo) battery accidentally falling into water. The total mass of the ship without battery and thruster is only $0.175\, \textrm{kg}$. The thruster is installed in the middle of the ship. It is fixed under the ship, about $0.5\, \textrm{cm}$ under the floats, by a movable piece made of polystyrene. This setup allows us to remove easily the thruster to rinse it and change the electrodes. Indeed, at high concentration and high electric current, the oxidation of aluminum occurs rapidly and electrodes can be damaged (see \ref{sec:fluid_electro}). To maximize the efficiency of the ship new electrodes and fully charged batteries were used before each experiment.

The Lorentz force propelling the ship is maximum when the current density and the local magnetic field are orthogonal to each other. It is thus interesting to control the electrical and magnetic field geometries to maintain this perpendicularity in a large fluid volume. With an external propulsion system, the magnetic field geometry is relatively complex, whereas internal conductive propulsion allows to impose this perpendicularity in large thruster volumes, thus maximizing the ship speed \cite{thibault1994status}. The thruster chosen in this paper is thus an internal conductive one, made of an electric and magnetic circuit disposed in a way to generate electric and magnetic fields orthogonal to each other, as detailed in section \ref{sec:elec}, \ref{sec:mag} and \ref{sec:thruster}. To avoid any unwanted propagation of the electric current in the magnetic bridge used to close the magnetic field (section \ref{sec:mag}), the electric circuit is isolated from the magnetic one using 3mm thick pieces of plastic (Fig~\ref{fig02}C). Electrodes are connected to the battery using a common XT60 plug widely used in scale modeling. Note that, to facilitate the connection between electrodes and cables, part of the electrodes are outside the thruster and isolated using specific electrical tape. It is important to isolate this part to avoid any unwanted electric current to propagate. In fact, outside the thruster, the magnetic field is opposite to the one inside which mean that if some electric current can propagate between electrodes at that point, the resulting Lorentz force will be in the opposite direction to the one generating the motion of the ship and hence will dramatically decrease the ship speed. 
 
Concerning the thruster orientation under the ship, two different configurations are possible by using either a vertical magnetic field (aligned with gravity) with a horizontal current or the opposite. The advantage of using a horizontal magnetic field is that the ship is then automatically guided by the Earth magnetic field, which is mostly horizontal when being far from the poles. The magnetic circuit of the ship thus behaves as a compass needle. On the other hand, to drive the ship in an arbitrary direction, a vertical magnetic field and horizontal current is a better choice. Note that using a horizontal current could also be beneficial for the electric circuit since it tends to avoid accumulation of rising electrolysis bubbles on the top electrode \cite{tempelmeyer1990electrical}. In this work, we choose to consider a horizontal magnetic field because the velocity is easier to measure with a guided ship (a mechanical guide would generate unwanted friction), and the tank is thus oriented along the magnetic East-West line, obtained from a compass such that the magnetic circuit is along a North-South line.

{Two kinds of experiments are performed for each battery and salt concentration. The first one consists of the so-called bollard pull experiments where the boat is directly attached to a dynamometer to measure the generated traction force (section \ref{sec:magGG3}). The second is a more common velocity measurement (section \ref{sec:thruster}) where the motion of the boat, in front of a ruler, is recorded at 60 frame per second. Then, by following a marker on the boat using an image processing software (ImageJ \cite{imageJ2012}), we extracted its displacement as a function of time, allowing us to calculate its velocity. Note that we carefully ensure that the boat travels as straight as possible during the experiments.} All the experiments have been performed using the same thruster made of two Neodymium N40 magnets representing the best compromise between weight and generated magnetic field's strength. Fig~\ref{fig01} shows a picture of the experiment using the LiPo 6s battery. 
A summary of the different experimental parameters and their value as well as those used for the theory is presented in Table~\ref{tab1}.

\begin{table}[!ht]
\begin{adjustwidth}{-2.25in}{0in} 
\centering
\caption{
{\bf Experimental and theoretical parameters}}
\begin{tabular}{l | l | l}
\thickhline
Measurements\\
\thickhline
Magnets size &$L_x \times L_y  \times L_z$ & $88 \times 24 \times 10$ (mm)   \\
\hline
Electrodes size & $l_x \times l_y  \times l_z$  & $90 \times 1 \times 18$ (mm) \\
\hline
Floats size & $\mathcal{L}_x \times \mathcal{L}_z  $ & $500 \times 50 $ (mm) \\
\hline
Immersion depth$^a$ & $\mathcal{L}_x^*$ & $450$ mm  \\
\hline
Magnet residual field  & $B_r$ & $1.26$ T   \\
\hline
Inter-electrodes gap & $H$ & $14$ mm   \\
\hline
Inter-magnets gap & $W$  & $24$ mm   \\
\hline
Thickness of insulation & $\delta$ & $3$ mm   \\
\hline
Thruster fluid section & $S_d=H l_z$ & $252$  mm$^2$  \\
\hline
Thruster solid section$^b$ & $S_{th}$ & $1104\, \textrm{mm}^2 $ \\
\hline
Thruster fixing size & $D_y$  & $28$ mm  \\
\hline
Magnetic bridge thickness & $\delta$  & $3$ mm  \\
\hline
Kinematic viscosity & $\nu$  & $10^{-6}$ m$^2$.s$^{-1}$  \\
\hline
Water density & $\rho_0$  & $10^{3}$  kg.m$^{-3}$ \\
\hline
Ship mass & $m_{ship}$  & $0.715$ kg   \\
\hline
Battery$^c$ (LiPo) 3s  & mass of & $U_0=12.6$ V,\\
& $0.193$ kg & $I_{m}=55$ A\\
\hline
Battery (LiPo) 4s & mass of & $U_0=16.8$ V,  \\
& $0.203$ kg & $I_{m}=108$ A   \\
\hline
Battery (LiPo) 6s & mass of & $U_0=25.2$ V \\
 & $0.285$ kg  &  $I_{m}=135$ A  \\
\hline
LiPo internal resistance & $r_i$ & $0 \, \Omega $   \\
\hline
Fields ratio$^d$  & $f=B_r/B_0$ & $3.54$  \\
Magnetic bridge gain$^e$  & $\mathcal{G}$ & $1.22$\\
Thruster field$^f$ & $B$ & $303$ mT \\
Fringing ratio (Eq~\ref{eq:fri}) & $l_x^*/l_x$ & $1.26$ \\
\thickhline
Theory$^g$ \\
\thickhline
Energy coefficients & $\alpha_d=\alpha_{\infty}$ & $1$   \\
Momentum coefficients &$\beta_d=\beta_{\infty}$ & $1$   \\
 Mean sinus$^h$ of $\theta_2$ &$\overline{\sin \theta_2}$ & $1$   \\
Head loss coefficient$^i$ & $\xi$ & $1.78$   \\
Darcy friction factor & $f_D$ & $0.3164 \, Re^{-1/4}$   \\
Form drag coefficient$^j$  & $C_d^f$ & $1$\\
Wave drag coefficient$^k$ & $C_d^w$ & $0$ \\

\end{tabular}
\label{tab1}
\begin{flushleft} 
$^a$ $\mathcal{L}_x^*$ is the length of the rectangle of the same area and width $\mathcal{L}_z $ than one float (in the plane xOz parallel to the water surface)\\
$^b$ i.e. the thruster section involved in the form drag\\
$^c$ $I_{m}$ is the maximum current sustainable by the battery.\\
$^d$ with $B_0=356\, \textrm{mT}$ the mean field at the magnet's pole surface.\\
$^e$ equations (\ref{eq:gain}) and (\ref{eq:wGain}), where  $B_{sat} = 2\, \textrm{T}$  (section \ref{sec:mag}). \\
$^f$ including the magnetic bridge gain of $22 \%$. \\
$^g$ values chosen for our theoretical calculations. \\
$^h$ $\overline{\sin \theta_2}= \int_{\mathcal{V}} ||\boldsymbol{j} \times \boldsymbol{b}||/(jb) \, \textrm{d}\mathcal{V}$, with the thruster volume $\mathcal{V}=l_x l_z H$. \\
$^i$ total singular head loss coefficient of the thruster. \\
$^j$ when $Re >10^3$, $C_d^f \approx 1$ for an immersed circular disk, as well as for immersed 2D wedges with a half-vertex angle of $27^{\circ}$, which corresponds to the (plumb) stem angle of our hulls (wave drag and the skin friction drag are negligible here). \\
$^k$ $C_d^w=0$ in our experiment (shallow water, $Fr<1$). 
\end{flushleft}
\end{adjustwidth}
\end{table}

\section{Thruster electrical properties}\label{sec:elec}

\subsection{Electrical circuit} \label{sec:elleC}

The electrical circuit consists of a voltage source (LiPo batteries) of internal resistance $r_i$, maintaining an electric potential $U_0$ between two rectangular electrodes of size $l_x \times l_y \times l_z$ separated by a distance $H$ in the $y$-direction. These two electrodes generate an electric current in the fluid of conductivity $\sigma$ (the fluid being salt water in our case). In presence of a magnetic field $\boldsymbol{b}$ and assuming that the fluid is flowing at a velocity $\boldsymbol{u_d}$ inside the thruster, the current density can be described using the local Ohm's law (e.g. \cite{Mitchell1988} or \cite{Boissonneau1999b} on MHD propulsion)
\begin{eqnarray}
\boldsymbol{j}=\sigma(\boldsymbol{E}+\boldsymbol{u_d} \times \boldsymbol{b}), \label{eq:Ohm1}
\end{eqnarray}
where $ \boldsymbol{j}$ is the current density and $ \boldsymbol{E}$ the electric field. Assuming uniform fields, the rectangular geometry allows the integration of Eq~(\ref{eq:Ohm1}), leading to 
\begin{equation}
I = l_x l_z\sigma \left(\frac{U}{H}+ k u_d B \right),
\end{equation}
using $|\boldsymbol{E}|= U /H$, where $U$ is the electrical potential existing in the fluid. This leads to the global Ohm's law
\begin{equation}
U =  (r+r_i) I+ k\, u_d BH,  \label{eq:Ohm1bis}
\end{equation}
where $ r=H/(\sigma l_x l_z)$ is the fluid electrical resistance in absence of fluid flow, $B$ the volume-averaged magnetic field strength, and $k$ the volume-averaged sinus of the angle between $\boldsymbol{u_d}$ and $\boldsymbol{b}$. Note that $U$ may differ from the potential $U_0$ imposed by a voltage source because of the fluid electrolysis chemical reactions, this will be presented in section \ref{sec:fluid_electro}.

The evolution of the fluid conductivity $\sigma$ as a function of salt concentration $C$ can be described using the Kohlrausch law, also called Debye-H\"uckel-Onsager equation (see \cite{wright2007introduction} for a theoretical derivation of the law). This law reads
\begin{eqnarray}
\sigma=a_0 C-b_0\, C^{3/2} , \label{eq:Kohl}
\end{eqnarray}
which gives $\sigma=0$ for $C=0$ and  $C_0=(a_0/b_0)^2$. Eq~(\ref{eq:Kohl}) also gives a maximum conductivity $\sigma_{max}=4 a_0^3/(27 b_0^2)$ at the concentration
\begin{eqnarray}
C_{max}=\frac{4}{9} \frac{a_0^2}{b_0^2}. \label{eq:Copt0}
\end{eqnarray}
The parameters $a_0$ and $b_0$ can be estimated with theoretical expressions \cite{wright2007introduction}, but, to be as close as possible from actual measurements, we choose here to fit tabulated values of the literature for a NaCl electrolyte \cite{JR9370000574}. At $20^{\circ}\, \textrm{C}$, this gives $a_0 \approx 2071\cdot 10^{-4}\, \textrm{S.m}^{2}.\textrm{kg}^{-1}$ and  $b_0 \approx 98.32 \cdot 10^{-4}\, \textrm{S.m}^{7/2}.\textrm{kg}^{-3/2}$, leading to the maximum conductivity  $\sigma_{max} \approx 13.6\, \textrm{S.m}^{-1}$ for $C_{max} \approx 197\, \textrm{kg.m}^{-3}$. Note that $C_0 \approx 444\, \textrm{kg.m}^{-3}$ is far larger than the solubility of NaCl in water at $20^{\circ}\, \textrm{C}$, and cannot thus be reached.

Experiments in salt water show that two supplementary effects have to be taken into account in Eq~(\ref{eq:Ohm1bis}). The fringing effects of the electrodes and the fluid electrolysis. 
When a voltage difference is applied between two conductive objects (the electrodes here) the generated electric field extends over a distance larger than the electrode itself. This is called the fringing effect, it means that the electric current actually flows on a length $l_x^*$ larger than $l_x$, leading to a smaller fluid electrical resistance $r^*<r$. 
In our configuration, the fringing effect can be estimated analytically using equation (10) of \cite{leus2004fringing}, giving $l_x^*= \varpi l_x$, with
\begin{eqnarray}
\varpi=1+\frac{H}{\pi l_x} \left[ 1+ \ln \left( \frac{2 \pi l_x}{H} \right) + \ln \left( 1+ 2 \zeta + 2 \sqrt{\zeta+\zeta^2} \right) \right],  \label{eq:fri}
\end{eqnarray}
where $\zeta=l_y/H$, and $l_y$ is the electrode thickness. The actual fluid resistance (without fluid flow) is thus rather $r^*=H/(\sigma l_x^* l_z)$.

\subsection{Fluid electrolysis} \label{sec:fluid_electro}
Imposing a voltage difference between two electrodes in an electrolyte (e.g. salt water) leads to fluid electrolysis, i.e. non-spontaneous electrochemical reactions driven by an electric current. In a NaCl solution, we can consider the water reduction to hydroxide and hydrogen gas at the cathode (see \cite{bennett1978site,leroy1978time} for details)
\begin{eqnarray}
\textrm{H}_2 \textrm{O}(l) +2 \textrm{e}^- \rightarrow \textrm{H}_2(g) + 2\textrm{OH}^-(aq),
\end{eqnarray}
and oxidation of chloride to chlorine 
\begin{eqnarray}
2\textrm{Cl}^-(aq) \rightarrow \textrm{Cl}_2(g) +2 \textrm{e}^-
\end{eqnarray}
at the anode. The overall electrolysis of aqueous NaCl results in hydrogen and chlorine gas formation and can be written as
\begin{align}
2\, \textrm{NaCl}(aq) + & 2\, \textrm{H2O}(l) \rightarrow \\
& 2\, \textrm{Na}^+(aq) + 2\, \textrm{OH}^-(aq) + \textrm{Cl}_2(g) + \textrm{H}_2(g). \nonumber
\end{align}
Note that the standard potential of $\textrm{Na}^+$ reduction is $\textrm{E}^0=-2.71 \textrm{V}$ whereas the one for reduction of water is $\textrm{E}^0=-1.23 \textrm{V}$ which means that, in aqueous solutions, water reduction will prevail. Chlorine gas $\textrm{Cl}_2$ will rapidly be dissolved in water (giving $\textrm{ClO}^-_3$ ions, see e.g. \cite{Boissonneau1999}) meaning that hydrogen gas $H_2$ would mainly be observed in this case. Depending on the electrode material, reactions implying the electrodes can also occur. For instance, aluminum electrodes can be damaged by oxidation \cite{tempelmeyer1990electrical}
\begin{eqnarray}
\textrm{Al}^{3+}(aq)+ 3 \textrm{e}^-  \rightarrow \textrm{Al}(s) ,
\end{eqnarray}
with a standard potential of $-1.66\, \textrm{V}$. Note that stainless electrodes can be used, but copper electrode should be avoided because an oxyde or chloride film is formed on the copper anode limiting the current \cite{tempelmeyer1990electrical}

Having the lowest standard potential, the water oxidation prevails onto other reactions, and the electrolysis starts as soon as the voltage is larger than $1.23\, \textrm{V}$. However, any voltage larger than $2.71\, \textrm{V}$ will drive the three reactions. Note that the production of hydroxide $\textrm{OH}^-$(aq) at the cathode during the electrolysis will drastically increase the basicity of the solution. 

Over-potentials at the electrodes, due to electrolysis driven chemical reactions, can be estimated with the Butler-Volmer equation \cite{atkins2006atkins}. Two limiting cases of this equation appear in low and high over-potential regions. The high limit gives the so-called Tafel equation where the over-potential $\delta U$ is given by 
\begin{eqnarray}
\delta U=A_0 \ln \frac{j}{j_0}, \label{eq:Tafel}
\end{eqnarray}
where  $A_0$ is the so-called Tafel slope, $j$ is the current density, and $j_0$ is the so-called exchange current density \cite{atkins2006atkins}. Using $j\approx I/(l_x l_z)$, Eq~(\ref{eq:Tafel}) can be written under its usual form
\begin{eqnarray}
\delta U=E_0+A_0 \ln I, \label{eq:Tafel2}
\end{eqnarray}
where $E_0=-A_0\ln j_0$ and $A_0$ are two constants related to the reactions involved during the electrolysis. As discussed previously, to start the electrolysis of NaCl electrolytes, the voltage has to be larger than $1.23\, \textrm{V}$, and one can thus expect $E_0 \approx 1.23\, \textrm{V}$ \cite{tempelmeyer1990electrical}. The Tafel slope associated with the water oxidation is $0.3$ according to \cite{Convert1995}, and we thus expect $A_0=0.3$  (e.g. \cite{Convert1995}) as a typical value. This value also agrees with the detailed over-potential measurements of \cite{petrick1992results}, who note that the Tafel slope $A_0$ decreases as the conductivity increases. Thus, for a given current density, lower over-potentials are expected with higher conductivity solution. 

Finally, the total electrical potential $U_0$ is the sum of the over-potential $\delta U$ and the electrical potential $U$ in the fluid. Using equations (\ref{eq:Ohm1bis}) and (\ref{eq:Tafel2}), and taking fringing effect into account, it leads to
\begin{eqnarray}
    U_0 =\delta U+U= E_0+A_0 \ln I+RI+k u_d BH , \label{eq:elic}
\end{eqnarray}
where $R=r_i+r^*=r_i+H/(\sigma l_x^* l_z)$ and $l_x^*$ is given by Eq~(\ref{eq:fri}). Note that Eq~(\ref{eq:elic}) can be solved for the current, leading to
\begin{eqnarray}
    \ln I & =& \frac{V}{A_0}- \textrm{LambertW} \left(\frac{R}{A_0}\, \textrm{e}^{V/A_0} \right) , \label{eq:Lambert}
\end{eqnarray}
with $\textrm{LambertW}$ the Lambert W function and the voltage $V=U_0-E_0-k u_d BH$, where $V>0$ for a propulsive thruster. 

For very large currents, the rate of electrolysis driven reactions is large, and it may be questioned if this could reduce the local ions concentrations, which would limit the current. This phenomenon is actually quite common in electrolysis, leading to a limiting current density, reached when the electrolysis have consummated all the reactant present in the thin diffusion layer in contact with the electrodes. It can also be noticed that, in presence of a magnetic field as our case, the ions transporting the electric current are deviated by the Lorentz force. The use of a scalar conductivity $\sigma$, hidden in the scalar total resistance $R$ in Eq~(\ref{eq:elic}), which is considered as independent of the magnetic field, can then be a priori questioned. Finally, a strong flow may influence the electrolysis reactions, which also questions the validity of the over-potential terms in Eq~(\ref{eq:elic}). These three important questions are discussed in  S1 Appendix, where it is shown that Eq~(\ref{eq:elic}) remains valid in the usual ranges of parameters, i.e. for our small scale ship models as well as large scale MHD ships.

\subsection{Measurements of the electrical properties}\label{expe_elec}

\begin{figure}[h!]
\centering
\includegraphics[width=10cm]{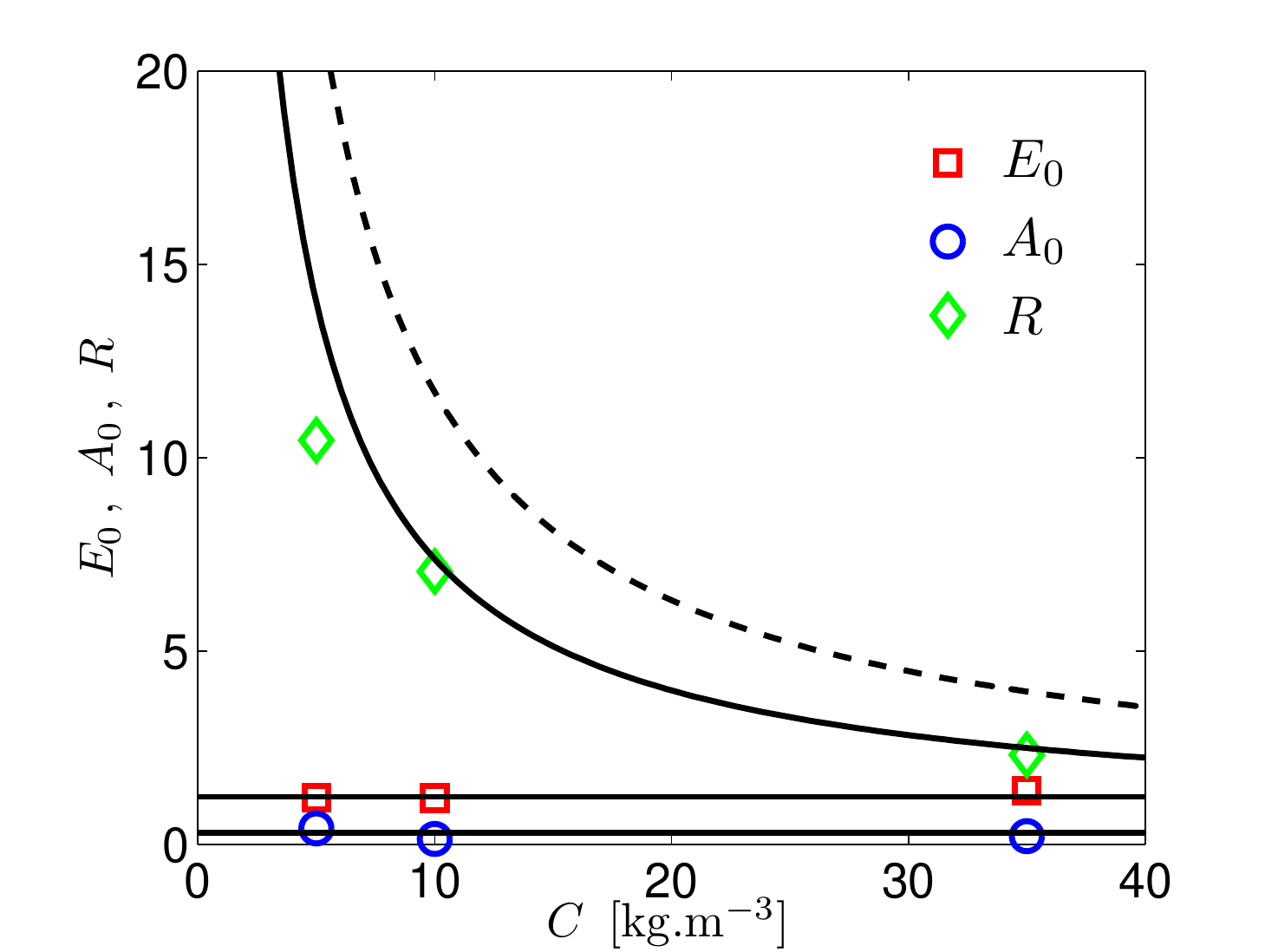} 
  \caption{{\bf Evolution of $E_0$, $A_0$ and $R$ with the NaCl concentration $C$}. The solid lines correspond to expected values (section \ref{sec:elec}), i.e. $A_0=0.3\, \textrm{V}$ (lowest one), $E_0=1.23\, \textrm{V}$ (intermediate one) and $R=r^*$ (the uppermost one). The dashed line shows the resistance $r$, i.e. the resistance without taking fringing effect into account. }
         \label{fig03}
\end{figure}

To test the electrical properties of the thruster, a conductometric cell made of two aluminum plates has been built. The electrode size is $2 \times 1.7\, \textrm{cm}$, with a thickness of $1\, \textrm{mm}$ separated one to another by a distance of $7\, \textrm{mm}$. Using a DC current generator, the current $I$ has been measured for $10$ voltages $U_0$ in the range $0.5-10\, \textrm{V}$, and for various concentrations of salt in the range $5-35\, \textrm{kg.m}^{-3}$. For a given concentration, we have checked that our data can successfully be fitted with a simplified version of equation \eqref{eq:elic}, where the magnetic term has been removed, i.e. with a function of the form $U_0=E_0+A_0 \ln I+R I$. These fits give the values of $E_0$, $A_0$ and $R$ shown in Fig~\ref{fig03}. The data are in excellent agreement with the expected theoretical values. Moreover, Fig~\ref{fig03} shows also that, taking into account fringing effect on the electrodes considerably improve the predictions of the fluid's electrical properties with salt concentration.  One can thus conclude that the theory predicts correctly the electrical behavior of the thruster in the ranges considered here, without any adjustable parameter.

Since fringing effects can be predicted by equation \eqref{eq:fri}, equation \eqref{eq:elic} can be used to obtain the evolution of the conductivity $\sigma$ from the values of $U_0$ and $I$. However the term $A_0\ln I$ makes analytical solution of equation \eqref{eq:elic} rather complicated. A simplification can be done using asymptotic behaviors of \eqref{eq:elic}. In the limit of zero current, the flow velocity inside the thruster is zero and thus \eqref{eq:elic} reduces to $U_0 = E_0$, on the opposite, for large currents, $RI \gg A_0 \ln I$ and \eqref{eq:elic} can be approximated by $U_0 \simeq RI + ku_dBH$. Combining these two asymptotic expressions leads to a simplified version of \eqref{eq:elic}

\begin{equation}\label{eq:14_simplified}
U_0 = E_0 + RI + ku_dBH,
\end{equation}
valid in both limits of small and large currents.

\begin{figure}[h!]
\centering
\includegraphics[width=10cm]{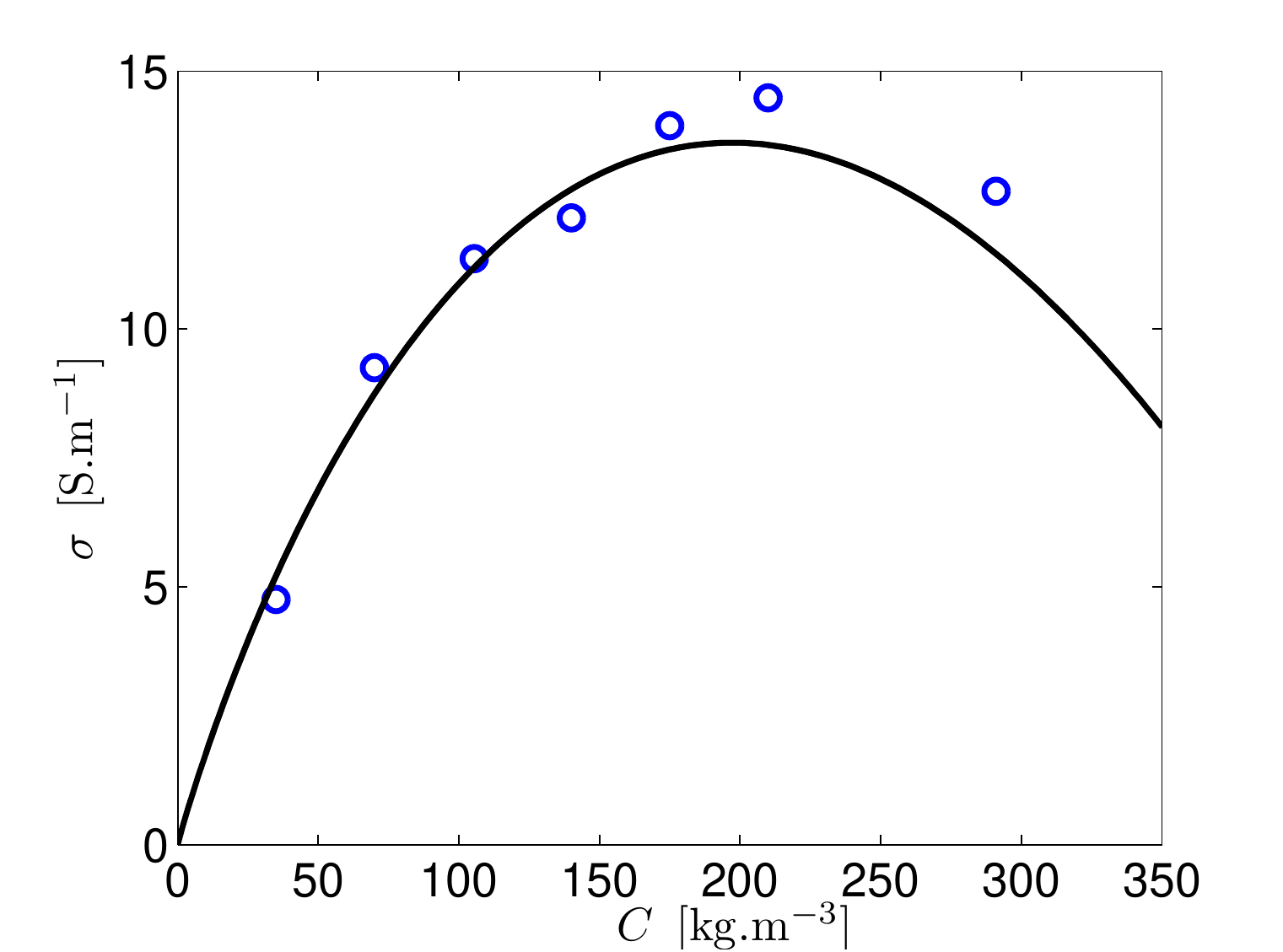} 
  \caption{{\bf Evolution of the conductivity with the NaCl concentration}. Our data (circles) are in excellent agreement with the Kohlrausch law (solid line), given by Eq~(\ref{eq:Kohl}), using  ($a_0 \approx 2071\cdot 10^{-4}\, \textrm{S.m}^{2}.\textrm{kg}^{-1}$,  $b_0 \approx 98.32 \cdot 10^{-4}\, \textrm{S.m}^{7/2}.\textrm{kg}^{-3/2}$). Note that the NaCl solubility in water is $360\, \textrm{kg.m}^{-3}$ at $25^{\circ}\, \textrm{C}$.}
         \label{fig04}
\end{figure}

Given the difficulty to measure $(U_0,I)$ on a moving ship, a static ship has been first considered. Indeed, since the term $k u_d BH$ is negligible in equation \eqref{eq:14_simplified}, these values should not change appreciably for a moving ship. 
Data are fitted with equation \eqref{eq:14_simplified}, where $k u_d BH$ has been neglected, to provide $R$. Then, using Table~\ref{tab1} and Eq~(\ref{eq:fri}), the conductivity $\sigma$ can be estimated for each salt concentration. The results, shown in Fig~\ref{fig04}, are in good agreement with the Kohlrausch law (\ref{eq:Kohl}), obtained from the tabulated values of the literature. All the different electrical measurements are in good agreement with the expected values from the theory, confirming the validity of equation \eqref{eq:elic}.

\section{Thruster magnetic properties} \label{sec:mag}

\subsection{Magnetic circuit} \label{sec:magGG}

The magnetic circuit consists of two neodymium N40 cuboidal magnets, of size $L_x \times L_y \times L_z$, separated by a distance $W+L_z$ in the $z$-direction (such that $W$ is the distance between their surface) and magnetized along $z$ with a residual magnetic field density $B_r$. The generated magnetic field is then channelized by placing the magnets on a high magnetic permeability (ferromagnetic) U shaped piece of iron which will be called magnetic bridge in the rest of the paper. The magnetic bridge is to magnetic circuit what electrical wires are to electric one; it allows to close the magnetic field. The magnetic field generated by these two cuboidal magnets can be evaluated using the analytical expression of \cite{camacho2013alternative}, based on \cite{yang1990potential}. Note that the results fully agree with \cite{engel2005calculation} provided that the magnetization field $B_r \boldsymbol{e_z}$ is added within the magnet.
The obtained results allow the calculation of the volume-averaged magnetic field $B$ required in section \ref{sec:elec}.
The use of a magnetic bridge is helpful for two main reasons. First, from an experimental point of view, it forces the magnetic field to flow through the bridge, reducing the generation of unwanted forces between the magnetic circuit and the ferromagnetic materials in the surrounding environment. Second, it slightly increases the magnetic flux density by reducing the magnetic reluctance of the circuit.

This small increase of magnetic flux due to the magnetic bridge can be estimated with simple arguments. Given the problem symmetry, we only consider for this estimation a single magnet with a magnetic bridge. The magnetic flux circulating through the magnet (of reluctance $\mathcal{R}_m=L_z/(\mu_0 L_x L_y)$) can flow through different media following two different main paths. It can either flow in the magnetic bridge (of reluctance $\mathcal{R}_i$) then through the gap between the iron bridge and the magnet (of reluctance $\mathcal{R}_g=W/(\mu_0 L_x L_y)$), or flow only in the surrounding medium (of reluctance $\mathcal{R}_a$), which can be seen as the salt water reluctance in the absence of iron bridge. This 'choice' of two possible ways for the magnetic flux leads to the association in parallel of $\mathcal{R}_a$ and $\mathcal{R}_g+\mathcal{R}_i$ for the total water reluctance, in series with $\mathcal{R}_m$, which gives
\begin{eqnarray}
\mathcal{R}_{tot}=\mathcal{R}_m+\frac{\mathcal{R}_a\, (\mathcal{R}_g+\mathcal{R}_i)}{\mathcal{R}_a+ \mathcal{R}_g+\mathcal{R}_i}. \label{eq:GGay}
\end{eqnarray}
Since the magnetic bridge has a high magnetic permeability, $\mathcal{R}_i$ is a priori negligible. 
In the limit of very small gap, $ \mathcal{R}_{tot} \approx \mathcal{R}_m$, whereas in the limit of infinite gap, we recover $ \mathcal{R}_{tot} \approx \mathcal{R}_m+\mathcal{R}_a$. In the first case, the magnet is 'shunted', and generates the magnetic flux density $B_r$, whereas the field of an isolated magnet is recovered in the second case. The magnetic flux is thus increased by the bridge of a factor $\mathcal{G}_{(\mathcal{R}_i=0)}$ given by
\begin{eqnarray}
\mathcal{G}_{(\mathcal{R}_i=0)}=\frac{\mathcal{R}_m+\mathcal{R}_a}{\mathcal{R}_{tot}}=\frac{1+\mathcal{R}_a/\mathcal{R}_m}{1+\frac{\mathcal{R}_a/\mathcal{R}_m}{1+\mathcal{R}_a/\mathcal{R}_g}}. \label{eq:factor}
\end{eqnarray}
Designating $f=B_r/B_0$ the ratio between the mean magnetic flux density generated by the shunted magnet (i.e. $B_r$) and the mean magnetic field generated by the isolated magnet (i.e. $B_0$), the magnetic flux conservation $(\mathcal{R}_m+\mathcal{R}_a)B_0=\mathcal{R}_m B_r$ gives $\mathcal{R}_a/\mathcal{R}_m=f-1$. Thus, noting $w=\mathcal{R}_g/\mathcal{R}_m$, we obtain $\mathcal{R}_a/\mathcal{R}_g=(f-1) /w$, and the magnetic flux amplification factor $\mathcal{G}_{(\mathcal{R}_i=0)}$ reduces to
\begin{eqnarray}
\mathcal{G}_{(\mathcal{R}_i=0)}=\frac{f (w+f-1)}{f(w+1)-1}, \label{eq:gain}
\end{eqnarray}
where  the factor $f$ depends on the magnet shape. However, magnets are commonly designed with the Evershed criterion, which prescribes $f=2$ for magnets with a linear demagnetization curve of the form $ B \approx \mu_0 \mathcal{H} + B_r$, where $\mathcal{H}$ is the magnetizing field. Most modern neodymium magnets present a demagnetization curve of this kind, and their recoil lines are thus very close to their demagnetization curve: these modern magnets operate thus along their demagnetization curve. Using the Evershed criterion, which maximizes the energy product $BH$ in the magnet by finding the rectangle of maximum area fitting below the demagnetization curve  $ B \approx \mu_0 \mathcal{H} + B_r$, one can obtain  that $B\mathcal{H}$ is maximum for $B=B_r/2$, i.e. for $f=2$. Indeed, using the equation $y=A_1 x+A_0$ for our linear demagnetization curve, the product $xy=x(A_1 x+A_0)$ is maximum for $(x,y)=(A_0/(2A_1),A_0/2)$, thus for $B=B_r/2$ if $A_0=B_r$ (a circular demagnetization curve of equation $y=\sqrt{Br^2-x^2}$ would maximize $xy$ for $y=B_r/\sqrt{2}$, which is close to $B_r/2$.). 

\begin{figure}[h!]
\centering
 \begin{tabular}{ccc}
\includegraphics[width=10cm]{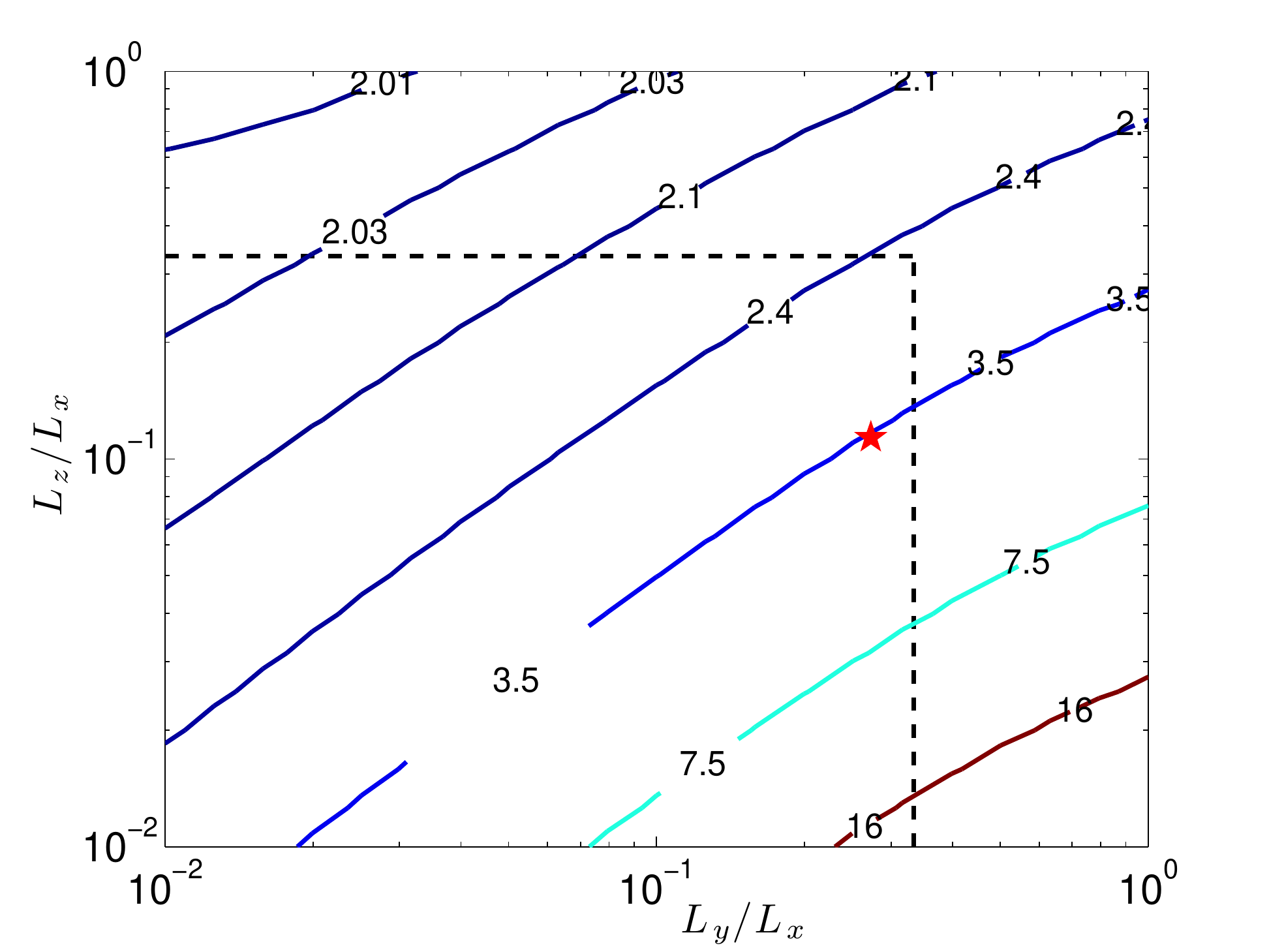} 
 \end{tabular}
  \caption{{\bf Contour lines of $f$ as a function of the cuboidal magnet dimensions.} The dashed black lines show $L_y/L_x=1/3$ and $L_z/L_x=1/3$, and the magnets used in this work are represented by the star ($Oz$ being the magnet magnetization direction and $Ox$ the mean flow velocity in the thruster).}
         \label{fig05}
\end{figure}

Using the formula of \cite{camacho2013alternative}, $f$ can be calculated for different magnet dimensions (Fig~\ref{fig05}). To channel the flow, the MHD thruster dimension in the flow direction $Ox$ is expected to be large compared to the two others in the normal directions ($Oy$ and $Oz$, the latter being the magnet magnetization direction). Choosing arbitrarily a thruster 3 times longer ($Lx$) than wide ($Lz$) or high ($Ly$), we only consider magnets with $L_y/L_x \leq 1/3$ and $L_z/L_x \leq 1/3$, i.e. magnets in the rectangle delimited by the dashed lines in Fig~\ref{fig05}. In this rectangle, the Evershed criterion shows that $L_z/L_x$ should be as large as possible and $L_y/L_x$ as small as possible. We thus choose magnets with quite large $L_z/L_x$, but also with a quite large $L_y/L_x$ because of the magnets availability. The magnets used in this study are represented by the red star in Fig~\ref{fig05}. {Note that a rough but simple analytical estimate of $f$ can be obtained for our magnets by estimating $B_0$ with equation (1) in S4 Appendix, i.e. with the axial field of a solenoid of radius $a$ and length $L_z$. At the surface of the magnet (i.e. for $z=L_z/2$), this provides $f \approx 2\sqrt{1+(a/L_z)^2}$. With the apparent radius $a \approx \min (L_x/2,L_y/2)$, we obtain $f \approx 3.1$, which is quite close to the actual value $f \approx 3.5$.}

As $w$ vanishes, the magnetic bridge increases the magnetic flux by a certain factor, given by Eq~(\ref{eq:gain}), between $1$ (isolated magnet, i.e no magnetic bridge) and $2$ (shunted magnet, i.e no air gap between the magnetic bridge and the magnet). It is important to note that the magnetic bridge has usually the same length $L_x$ as the magnet, but a smaller thickness $\delta$, i.e. a smaller cross section $\delta L_x$, which increases the magnetic field within the bridge (magnetic flux conservation). However, beyond a certain value $B_{sat}$ for the magnetic field, the ferromagnetic material constituting the magnetic bridge saturates and the magnetic field leaks then in the surrounding environment. To avoid that, the magnetic bridge thickness has to be large enough, i.e.
\begin{eqnarray}
\delta \geq \frac{B L_y}{B_{sat}}, \label{eq:ironSAT}
\end{eqnarray}
with $B_{sat} \sim 1-2\, \textrm{T}$ for most ferromagnetic alloys. In practice, using iron and neodymium magnets leads easily to saturation, and the gain $\mathcal{G}_{(\mathcal{R}_i=0)}$ is then reduced to a smaller gain $\mathcal{G}$ because of magnetic leaks. Relaxing  the hypothesis of negligible $\mathcal{R}_i$ assumed above, the magnetic flux conservation $ B_0 \mathcal{R}_m = B_{sat} \mathcal{R}_i$ gives $\mathcal{R}_i/\mathcal{R}_m = B_0 L_y/(\delta B_{sat})$. Using Eq~(\ref{eq:GGay}), we obtain that the gain  $\mathcal{G}=(\mathcal{R}_m+\mathcal{R}_a)/\mathcal{R}_{tot}$ is actually given by Eq~(\ref{eq:gain}) provided that $w$ is replaced by 
\begin{eqnarray}
w=\frac{ \mathcal{R}_g}{ \mathcal{R}_m}+\frac{ \mathcal{R}_i}{ \mathcal{R}_m}=\frac{W}{L_z}+\frac{L_y}{\delta} \label{eq:wGain} \frac{B_0}{B_{sat}}
\end{eqnarray}

\subsection{Simulations of the magnetic properties} \label{sec:magGG2}

Using the commercial software COMSOL (based on the finite elements method), we solved the magnetic flux conservation for a single magnet, two magnets, and for the total magnetic circuit (two magnets and the magnetic bridge). This allows us to check our different formula related to the magnetic circuit, especially the expression of the field generated by a cuboidal magnet (given by \cite{camacho2013alternative}). Our simulations also show that the magnetic bridge gives a gain of $60\%$ in absence of magnetic saturation, in good agreement with the value $\mathcal{G}_{(\mathcal{R}_i=0)} \approx 1.58$ given by  Eq~(\ref{eq:gain}). 

Then, using a Hirst GM05 Gaussmeter, we measured a magnetic field of $\approx 300\, \textrm{mT}$ at the distance $W/2$ of the surface of the magnet's north pole, in the middle of the thruster. With a single magnet, and in absence of magnetic bridge, COMSOL simulations result in a field of $B_z=124.26\, \textrm{mT}$, in excellent agreement with the field $B_z=124.04\, \textrm{mT}$ given by the formulas of \cite{camacho2013alternative}. One can deduce the experimental magnetic bridge amplification factor $\mathcal{G}=1.21$, in good agreement with our theoretical gain of $22 \%$.

\subsection{Measurements of the magnetic properties} \label{sec:magGG3}

\begin{figure}[h!]
\centering
\includegraphics[width=10cm]{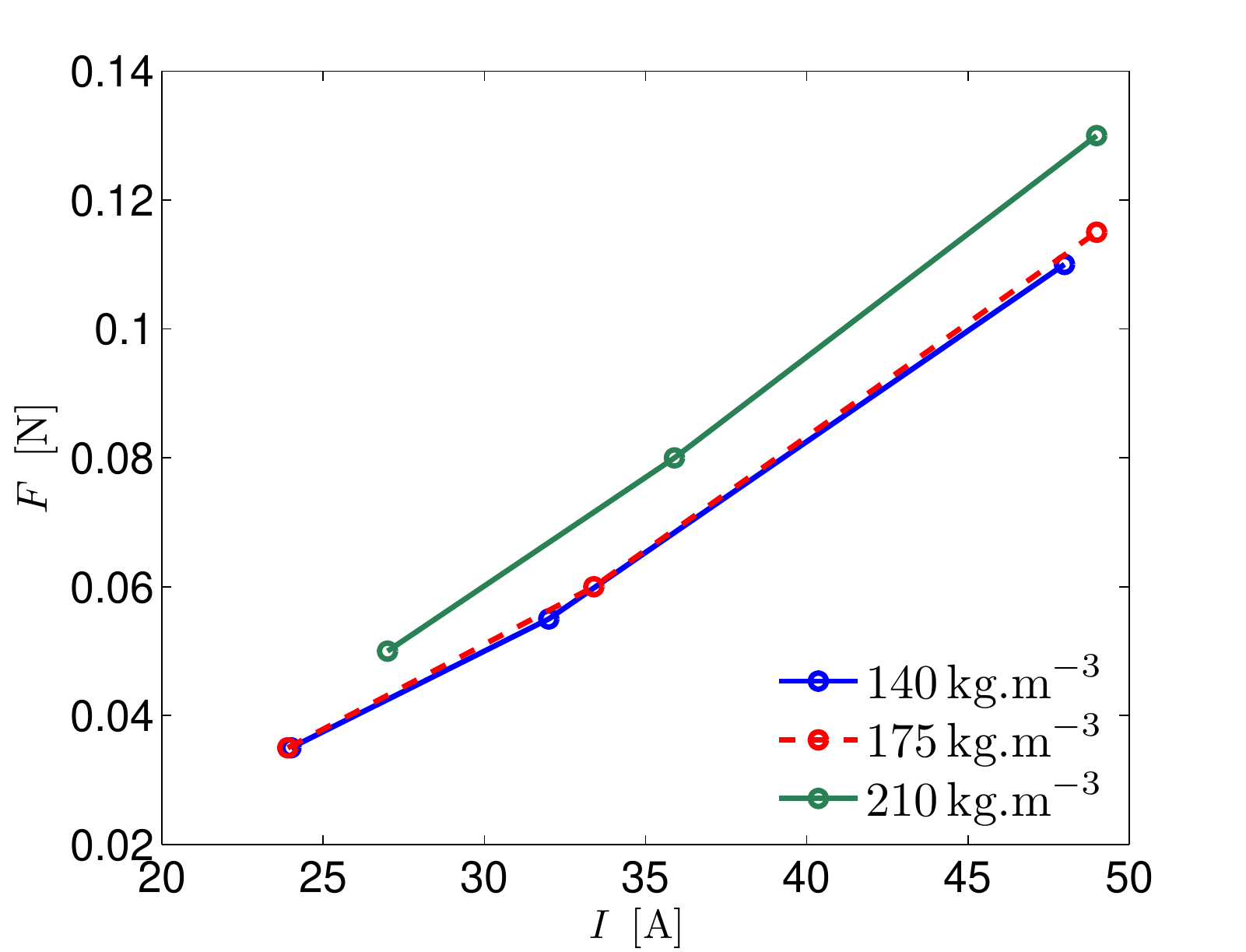} 
  \caption{{\bf Evolution of the traction force with the current intensity}. The theory predicts $F = IBH $, leading to a collapse of the data on a single line ($BH$ being constant here). Note that the fit of the data gives a non-zero intercept, probably due to friction in our dynamometers.}
         \label{fig06}
\end{figure}

To exploit further the static ship measurements used to determine the electrical properties of the thruster, we also attempted to measure the ship traction force $F$ with dynamometers (bollard pull tests). However, most of our measurements of $F$ were largely disturbed by radiated/reflected waves, residual currents, etc. Reliable results have only been obtained for the largest values of $C$ because $F$ becomes then significant compared to perturbations. These results are shown in Fig~\ref{fig06}. 

Since the theoretical force generated by the thruster corresponds simply to the Lorentz force, generated by electromagnetic fields, $F=IBH$, all the measured traction forces are expected to collapse on a straight line of slope $BH$. A linear fit of the data shows that the intercepts are actually non-zero and negative, indicating an internal friction of $0.04-0.05\, \textrm{N}$ in our dynamometer. By contrast, the slopes are quite close, in the range $ 3.2-3.6 \, \textrm{mN.A}^{-1}$, in satisfying agreement with the theoretical slope $BH \approx 4.2\, \textrm{mN.A}^{-1}$ (error of $20 \%$).\\

\section{Study of the MHD ship}\label{sec:thruster}

\subsection{Fluid velocities and thrust}  \label{sec:thruster2}

\begin{figure}[h]
\centering
 \begin{tabular}{ccc}
   \SetLabels
   \L\B (-0.01*0.45) {\rotatebox{90}{$S_{\infty}$}} \\
   \L\B (0.065*0.45) {\rotatebox{90}{$u_{\infty}$}} \\ 
   \L\B (0.355*0.455) {\rotatebox{90}{$S_d$}} \\
   \L\B (0.535*0.455) {\rotatebox{90}{$u_d$}} \\
   \L\B (0.5*0.845) $u_{\infty}$ \\
   \L\B (0.5*0.12) $u_{\infty}$ \\
   \L\B (0.95*0.3) {\rotatebox{90}{$S_{\textrm{tank}}>>S_{\infty}$}} \\
   \endSetLabels
  \strut \AffixLabels{\includegraphics[width=0.9\hsize]{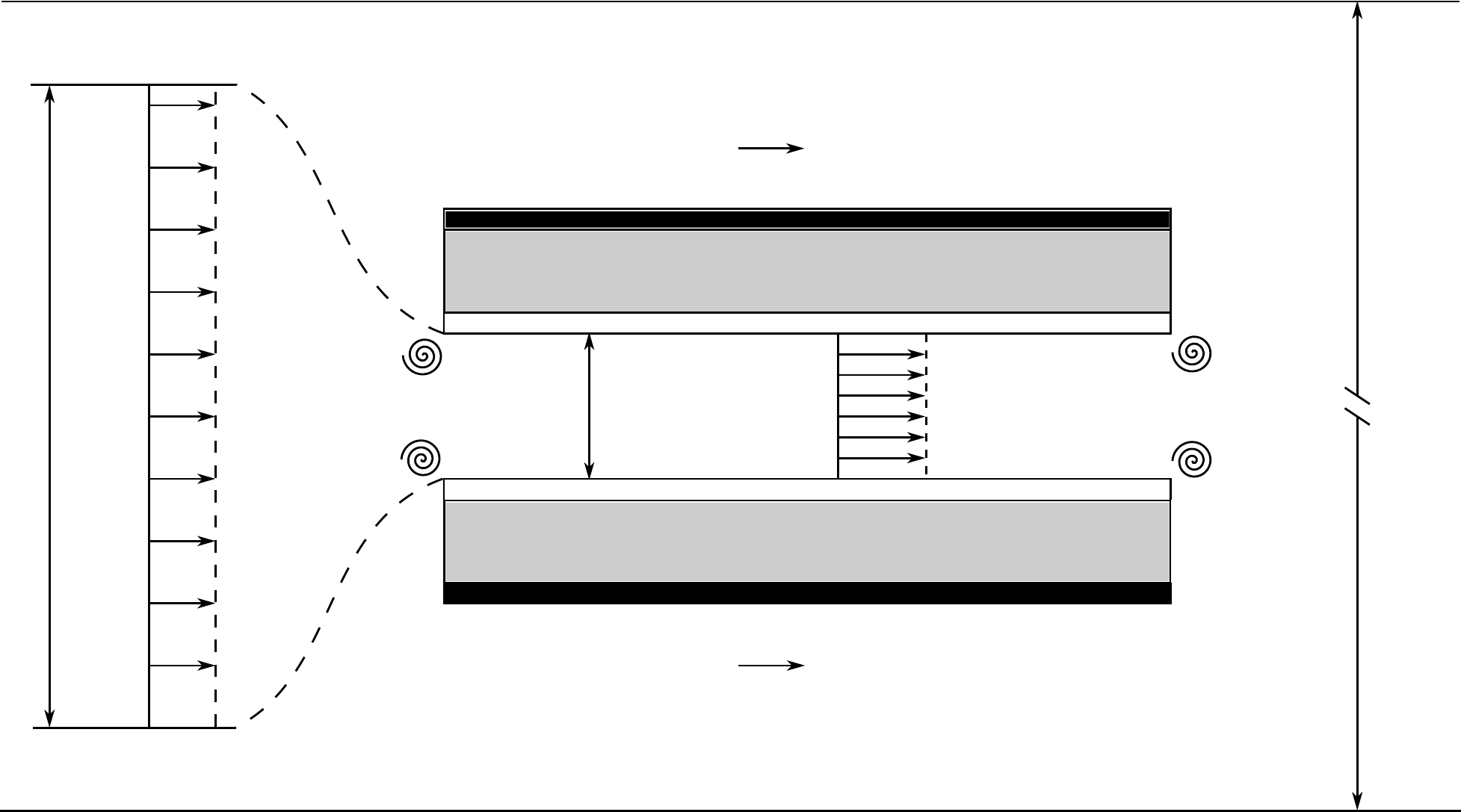}}
   \end{tabular}
  \caption{{\bf Sketch of the thruster and the different velocities taken into account in this study}. Four vortices are displayed to represent the singular head loss at the entrance and exit of the thruster.}
         \label{fig07}
\end{figure}
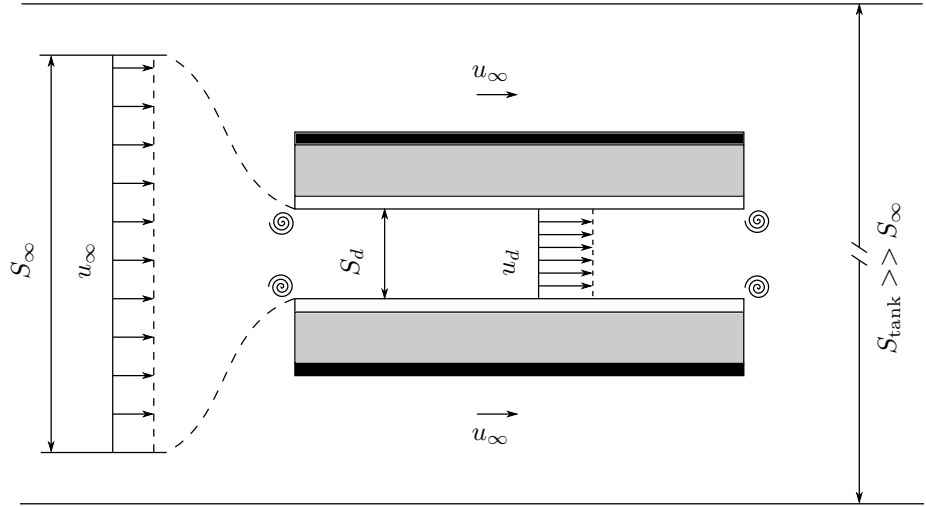

The MHD ship is propelled by a thruster which combine the electrical and magnetic circuits described in sections \ref{sec:elec} and \ref{sec:mag}, respectively. In the thruster, the electric current along $Oy$ and the magnetic field along $Oz$ generate a Lorentz force in the $x$-direction. This force generates a flow which propels the thruster, and thus the ship, in the opposite direction, at the velocity $- u_{\infty} \boldsymbol{e_x}$. In the frame moving with the thruster at $- u_{\infty} \boldsymbol{e_x}$, the mean flow velocity in the thruster is noted $u_d$. A schematic representation of the thruster and the different fluid velocities and cross section areas used for this study is shown in Fig~\ref{fig07}. Note that, contrary to the representation in Fig~\ref{fig07}, the present theoretical framework is developed for any kind of velocity profile inside the thruster. Nevertheless, as detailed below and in S2 Appendix, a uniform profile is a correct approximation for this configuration.

The cross section area in the thruster is constant, equal to $S_d=H l_z$. Since the fluid is assumed to be incompressible, volume conservation implies that the mean velocity $u_d$ in this section is also constant. Far from the thruster, the mean fluid velocity is $ u_{\infty}$. Using volume conservation, the section $S_{\infty}$ of fluid drawn in the thruster is thus given by
\begin{eqnarray}
 \frac{u_{\infty}}{u_d} = \frac{S_d}{S_{\infty}}  =  \lambda, \label{eq:vol}
\end{eqnarray}
where $\lambda$ is the velocity ratio (or the section areas ratio).

At steady state, the momentum conservation for the whole system (water and ship), reduces to the balance between the drag $F_D$ of the ship and the mean thrust $F$ of the thruster. The drag $F_D$  is  given by
\begin{eqnarray}
    F_D &=& \frac{1}{2} \rho S_w\,  C_d(u_{\infty})\,  u_{\infty}^2 , \label{eq:drgg}
\end{eqnarray}
where $\rho$ is the fluid density, $S_w$  the ship cross-section used for the ship drag, and $C_d(u_{\infty})$ the total drag coefficient of the ship. To estimate $C_d(u_{\infty})$, it is customary to write it as the sum of the skin friction, form drag and wave making drag coefficients
\begin{equation}
C_d(u_{\infty})=C_d^s+C_d^f+C_d^w.
\end{equation}
Note that $C_d^s$ is related to the force created by the friction between the fluid and the surfaces over which it is flowing. This skin friction drag force $F_D^s$ is usually estimated by considering the drag force generated by a fluid flowing over one side of a flat plate (parallel to the flow), which leads for instance to the Blasius law \cite{Schlichting},
\begin{eqnarray}
F_D^s=  \frac{1}{2} \rho S_{wet}\,  \frac{1.328}{\sqrt{Re_L}}\,  u_{\infty}^2 \label{eq:Blas}
\end{eqnarray}
where $S_{wet}$ is the total surface area of the plate in contact with the fluid, and $Re_L=u_{\infty} L / \nu$ the Reynolds number based on the length $L$ of the plate. 
Note that formula (\ref{eq:Blas}) is typically valid for $Re_l<5 \cdot 10^5$. Except for the internal wet surface of our propulsive thruster where the skin friction is taken into account through major head loss, our ship can be considered as a sum of flat plates of length $L_i$, leading to the total skin friction drag coefficient  in Eq~(\ref{eq:drgg}),
\begin{eqnarray}
    C_d^s=\frac{1.328}{S_w} \sum_i \frac{L_i \delta_i^*}{\sqrt{u_{\infty} L_i / \nu}}=\frac{1.328 \sqrt{\nu}}{S_w \sqrt{u_{\infty}  }} \sum_i  \delta_i^* \sqrt{L_i}, 
\end{eqnarray}
with $\delta_i^*=\delta_i \cos \theta_i$, where $\theta_i$ is the angle between the plate and $Ox$, and $L_i$ (resp. $\delta_i$) is the length of each plate along $Ox$ (resp. in the direction perpendicular to $Ox$). In our experimental setup, we have (in $\textrm{mm}^{3/2}$)
\begin{eqnarray}
\sum_i  \delta_i^* \sqrt{L_i} \approx 1670+126\, \mathcal{L}_y ,
\end{eqnarray}
where $\mathcal{L}_y$ is the immersion depth of the thruster top (in $\textrm{mm}$ in this formula). 

The coefficient $C_d^f$, related to the formation of a wake, depends on the exact shape of the hull and is typically of order $1$. Focusing on our experimental ship, one can confirm that $C_d^f \approx 1$. Indeed, $C_d^f \approx 1$ for an immersed circular disk at $Re >10^3$, i.e. for the cylindrical support of the thruster, and $C_d^f \approx 1$ for immersed 2D wedges with a half-vertex angle of $27^{\circ}$ (\cite{Hoerner1965a}), which corresponds to the (plumb) stem angle of our hulls.

The third coefficient $C_d^w$, related to gravity waves generation, is usually more difficult to estimate \cite{Tuck2002} and depends on the Froude number $Fr=u_{\infty}/\sqrt{gh}$, where $h$ is the water depth. In shallow water, it has been shown \cite{tuck1966shallow} that $C_d^w=0$ corresponding to our experimental regime $Fr<1$. 

The mean thrust $F$ is given by \cite{Boissonneau1999b}
\begin{eqnarray}
    F= \dot{m} (\beta_d u_d - \beta_{\infty} u_{\infty}) 
\end{eqnarray}
such that the final balance equation $F=F_D$ is 
\begin{eqnarray}
    \dot{m} (\beta_d u_d - \beta_{\infty} u_{\infty}) &=& \frac{1}{2} \rho S_w C_d(u_{\infty})  u_{\infty}^2 , \label{eq:BernoulliINF}
\end{eqnarray}
where  $\dot{m}=\rho S_d u_d=\rho S_{\infty} u_{\infty}$ is the mass flux through the thruster (mass conservation). Note that the velocity $u_{\infty}$ is not the inlet velocity, as ambiguously called by \cite{Mitchell1988}, because the inlet pressure is not a priori equal to the ambient pressure. As pointed out by \cite{Convert1995} and \cite{Gilbert1991a}, this point is the reason why the momentum balance has been sometimes erroneous in the literature anterior to the 90's (see also \cite{Brown1990,Davidson1992}).

In Eq~(\ref{eq:BernoulliINF}), $\beta_d$ (resp. $\beta_{\infty}$) is the momentum coefficient, or momentum correction factor, of the flow in the thruster (resp. of the upstream flow). For a given flow profile $u$ through a section S, this coefficient is defined by $\beta=1/S \cdot \int_S [u/\bar{u}]^2 \textrm{d} \tau$, where $\bar{u}$ is the mean flow. This correction factor is used to take into account non-uniform velocity profile when integrating the momentum equation. Typical values are $\beta=4/3$ (resp. $\beta=6/5$) for the Poiseuille flow in a cylindrical (resp. 2D plane) duct, and $\beta =1$ for a uniform velocity profile. However, the magnetic field will modify slightly the velocity profile in the thruster, leading to different values of $\beta$. To quantify this effect, one can consider the usual Hartmann flow between two planes. It shows that the magnetic field makes the flow closer to a uniform velocity profile, and thus $\beta$ is closer to $1$ (see S2 Appendix for calculation details). In any case, $\beta$ remains close to $1$, and we thus consider $\beta_d \approx \beta_{\infty}=1$, i.e. uniform flows, when comparing with our experimental results.

Substituting Eq~(\ref{eq:vol}) into (\ref{eq:BernoulliINF}) gives 
\begin{eqnarray}
 \frac{S_w C_d(\lambda u_d) }{2 S_d}   \lambda^2+  \beta_{\infty} \lambda -  \beta_{d}=0, \label{eq:lambd}
\end{eqnarray}
which has to be solved numerically for arbitrary $C_d(\lambda u_d)$. For the analytical calculations to be tractable, a constant $C_d$ can be assumed in Eq~(\ref{eq:lambd}), leading to
\begin{eqnarray}
\lambda= \frac{u_{\infty}}{u_d} = \frac{S_d}{S_{\infty}}  =  \frac{S_d}{S_w C_d} \left[ \sqrt{\beta_{\infty} ^2+\frac{2 \beta_d S_w C_d}{ S_d}} -\beta_{\infty}  \right], \label{eq:u_inf}
\end{eqnarray}
which relates the ship velocity $u_{\infty}$ and the thruster outflow mean velocity $u_d$ (in the frame of the thruster). In the limit $\beta_{\infty}^2 \ll 2 \beta_d S_w C_d /S_d$, Eq~(\ref{eq:u_inf}) shows that $\lambda \approx \sqrt{2 \beta_d S_d/(S_w C_d)}$. On the other hand in the limit where $\beta_{\infty}^2 \gg 2 \beta_d S_w C_d /S_d$, $\lambda$ tends to $\lambda=0$ (as $\lambda \approx \beta_d/\beta_{\infty}$ according to the next order). 

In order to close the system, another equation relating $u_{\infty}$ and $u_d$ is needed. Using the steady Navier-Stokes equation
\begin{eqnarray}
 \boldsymbol{u} \cdot \nabla \boldsymbol{u} = - \nabla p + \nu \nabla^2 \boldsymbol{u} + \boldsymbol{j} \times \boldsymbol{b}, \label{eq:NS}
\end{eqnarray}
where $\nu$ is the kinematic viscosity, $\boldsymbol{u}$ the velocity field, and $p$ the pressure, an average generalized Bernoulli equation can be derived when averaging over the thruster section $S_d$. This equation balances the flow kinetic energy gain through the thruster with the averaged total work of the Lorentz force and the total head loss $\chi$. It can be expressed as (e.g.\cite{Boissonneau1999b})
\begin{eqnarray}
  \frac{IBH\, \overline{\sin \theta_2}}{S_{d}} = \alpha_d \frac{1}{2} \rho    u_d^2 -   \alpha_{\infty}  \frac{1}{2} \rho   u_{\infty}^2 + \chi. \label{eq:mec2}
\end{eqnarray}
Where $\overline{\sin \theta_2}$ is the volume-averaged sinus of the angle between $\boldsymbol{j}$ and $\boldsymbol{b}$, and $\alpha_d$ (resp. $\alpha_{\infty}$) the energy coefficient, or kinetic energy correction factor, of the flow in the thruster  (resp. of the upstream flow). For a given flow profile $u$ through a section $S$, this coefficient is defined by $\alpha=1/S \cdot \int_S [u/\bar{u}]^3 \textrm{d} \tau$, where $\bar{u}$ is the mean flow velocity. As for the momentum correction factor $\beta$, $\alpha$ is used to take into account non-uniform velocity profiles when integrating the momentum equation. In the configuration studied here, $\alpha$ remains close to one (see S2 Appendix for details) and will be approximated to unity for the rest of the study.

The total head loss $\chi$ in the thruster consists of a linear head loss $\Lambda_1$ corresponding to the viscous friction of the fluid on the walls of the thruster, and  two singular head losses, $\Lambda_2$ and $\Lambda_3$, due to the thruster entrance and exit, respectively. We thus have
\begin{eqnarray}
\chi=\Lambda_1+\Lambda_2+\Lambda_3
\end{eqnarray}
The linear head loss, due to the relative velocity $u_d$ of the fluid with respect to the wall, is given by the Darcy Weisbach equation i.e. 
\begin{eqnarray}
\Lambda_1=f_D \frac{L_x}{D_h} \cdot \frac{1}{2} \rho u_d^2,
\end{eqnarray}
where $f_D$ is the Darcy friction factor, and $D_h$ the hydraulic diameter of the thruster. For a pipe, $D_h$ is simply the internal diameter, but for non-circular ducts, $D_h$ is rather given by the estimate $D_h=4A/P$, with $A$ the duct cross section, and $P$ the duct perimeter of cross section. For a rectangular duct, an even more accurate estimation is given by the Huebscher formula \cite{Huebscher1948},
\begin{eqnarray}
D_h=1.3\, \left[\frac{A^5}{(P/2)^2} \right]^{1/8}, \label{eq:Hueb}
\end{eqnarray}
which differs from equation $D_h=4A/P$ by $10 \%$.

Various expressions exist in the literature to estimate the Darcy friction factor $f_D$, depending on the Reynolds number $Re=u_d D_h/\nu$. For $Re < 2300$, $f_D$ is well estimated by the Hagen-Poiseuille law giving $f_D=64/Re$, whereas the Blasius estimate $f_D=0.3164 /Re^{1/4}$ is a good approximation for $4 000 < Re< 10^5$ in smooth pipes. Many other formulas exist, especially to take the pipe roughness into account. In any case, it is important to notice that the fluid velocity is required to calculate $f_D$, which introduces a supplementary non-linearity in the system. Note also that the dissipation, and thus $f_D$, is modified by the magnetic field. However, the Hartmann number $H_a$ is around $1$ in our experiment (see S2 Appendix for details), and the dependency of $f_D$ with the magnetic field can thus be neglected at leading order (see \cite{muller2002liquid} for quantitative estimates of this effect, e.g. their Fig~17).

Tabulated excess head coefficients or singular head loss coefficients, for pipe entrances and exits, are used to estimate the singular head loss $\Lambda_2+\Lambda_3$. These values have been obtained for a fluid at rest entering a pipe, or a pipe outflow in a tank of fluid at rest. To use these values, we have to consider the mean fluid velocity in the inertial frame of reference, where the fluid surrounding the thruster is at rest. In the inertial frame of reference, the mean flow velocity in the thruster is  $u_d-u_{\infty}$. Given that the singular head loss coefficient for an inward projecting/re-entrant (protruding pipe in a tank) is $0.78$, and the one for a pipe exit is $1$, we thus have 
\begin{eqnarray}
\Lambda_2+\Lambda_3=\xi \frac{1}{2} \rho (u_d-u_{\infty})^2=\xi \frac{1}{2} \rho (1-\lambda)^2 u_d^2, \label{eq:singH}
\end{eqnarray}
where $\xi=1.78$ is the sum of the  singular head loss coefficients. Note that the presence of other singular head losses would simply modify Eq~(\ref{eq:singH}). Here again, $\xi$ is a priori modified by the presence of a magnetic field, a effect which is neglected here because of the moderate values of the Hartmann number $H_a$ reached in our experiment (see S2 Appendix for details).

One can wonder if the head loss associated to the entrance and exit of the thruster could have been reduced using a different geometry in these zones. 
When a fluid exits a pipe into a much larger body of the same fluid, the velocity is reduced to zero and all of the kinetic energy is dissipated, thus the losses in the system are one velocity head, regardless of the exit geometry. However, the entrance loss coefficient can be made very small by using an appropriate rounded entrance geometry, which gives the lower bound $\xi \geq 1$. Note that the ship drag may be significantly increased by a different entrance geometry, reducing the overall interest of such a modification. 
{One may wonder if the electrolysis may modify the flow or the ship drag via the bubbles generation. According to \cite{Boissonneau1999}, the mean bubble diameter is $1-100\, \mu \mathrm{m}$ and the volume gas fraction is $10^{-4}-10^{-3}$ for seawater electrolysis in conditions close to our experimental setup. According to \cite{van2005drag}, we do not expect any influence of the bubbles on the drag, thus on the flow, which is confirmed by the very good agreement between our theory and our experimental results (see section \ref{sec:expeSpeed}).}

\subsection{Summary: thruster governing equations} \label{sec:finEQ}

As shown in the previous sections, the thruster dynamics can be described as follow: the imposed voltage generates an electric current, given by Eq~(\ref{eq:elic}), which generates a flow governed by Eq~(\ref{eq:mec2}), inducing an opposite electric current by a feedback term in Eq~(\ref{eq:elic}). Here, the magnetic field generated by this flow induced electric current is thus neglected. The validity of this approximation can be estimated using the magnetic Reynolds number $R_m=ud H/\nu_m$, with $\nu_m=(\sigma \mu)^{-1}$ the magnetic diffusivity and $\mu$ the fluid magnetic permeability. When $R_m \ll 1$, the induced field is negligible compared to the imposed one, and  the unknown current $I$ and velocity $u_d$ are then given by equations (\ref{eq:elic}), (\ref{eq:lambd}) and (\ref{eq:mec2}). Thus, the three unknowns $\lambda$, $u_d$ and $I$ are governed by   
\begin{eqnarray}
 \beta_{d} &=& \frac{S_w C_d(\lambda u_d) }{2 S_d}   \lambda^2+  \beta_{\infty} \lambda ,  \label{eq:new} \\
  U_0 & =& E_0+A_0 \ln I+RI+k u_d BH   \label{eq:elic22} \\
IBH &=& \mathcal{K}\, [1+\mathcal{G}(u_d) ]\, u_d^2 \label{eq:meca22} 
\end{eqnarray}
where
\begin{eqnarray}
\mathcal{K} &=& \frac{1}{2\,  \overline{\sin \theta_2}} \rho S_{d} [\alpha_d -   \lambda^2 \alpha_{\infty}  + \xi(1-\lambda)^2]   \label{eq:jjj},\\
\mathcal{G}(u_d) &=& f_D \frac{\rho S_d}{2\mathcal{K} \overline{\sin \theta_2}} \frac{L_x}{D_h} ,
\end{eqnarray}
allowing the calculation of $u_{\infty}$ and $S_{\infty}$ using Eq~(\ref{eq:vol}). For a static thruster, $u_{\infty}=0$, so $\lambda=0$, and the problem is then only governed by the two equations \eqref{eq:elic22} and \eqref{eq:meca22}. Note also that the inlet velocity (and pressure) is not involved in these equations, but can be calculated a posteriori using volume conservation and Bernoulli equation. 

Following \cite{thibault1994status}, it is of interest to estimate how the thruster electrical efficiency $\eta=\mathcal{P}_m/\mathcal{P}_e$ varies. Where $\mathcal{P}_m=IBH u_d$ is the mechanical power imparted on the fluid and $\mathcal{P}_e=U_0 I$ the electrical power given by the LiPo Battery. In the literature, the load factor
\begin{eqnarray}
K=\frac{U_0-E_0}{k u_d BH}, \label{eq:load0}
\end{eqnarray}
is often introduced \cite{weier2007flow}, which is the ratio between the effective voltage imposed to the fluid and the voltage induced by the flow. Using the load factor $K$, $\eta$ can conveniently be written as
\begin{eqnarray}
\eta=\frac{BH u_d}{U_0}=\frac{1}{k(K+K_0)}, \label{eq:effici0}
\end{eqnarray}
where $K_0=E_0/(k u_d BH)$ is usually neglected in the literature ($E_0 \ll U_0$). Since $K>1$ for a thruster, a good efficiency is reached for $K \geq 1$  \cite{Boissonneau1999b}. Considering the simpler Eq~(\ref{eq:14_simplified}), $K$ reduces to $K=1+RI/(k u_d BH)$, which gives
\begin{eqnarray}
\displaystyle
\eta =\left[k(1+K_0)+ \frac{R I}{u_d B H} \right]^{-1},  \label{eq:effici00}
\end{eqnarray}
i.e.
 \begin{eqnarray}
\eta = \left[k(1+K_0)+\frac{\mathcal{K} (1+\mathcal{G}(u_d) ) u_d}{B^2} \left( \frac{r_i}{H^2}+\frac{1}{\sigma \mathcal{V}} \right) \right]^{-1},  \label{eq:efficisi}
\end{eqnarray}
with $\mathcal{V}=l_x l_z H$ the volume of water in the thruster. Under the usual assumptions of the literature ($K_0=0$, $k=1$, $\mathcal{G}=0$, $r_i=0$), Eq~(\ref{eq:efficisi}) reduces to 
\begin{eqnarray}
\eta=\frac{1}{1+\mathcal{K} u_d/(\sigma \mathcal{V} B^2)} , \label{eq:etaS}
\end{eqnarray}
which shows that, for a given velocity $u_d$, the efficiency approaches $1$ when $\mathcal{V} B^2$ is increased \cite{weier2007flow}. Maximizing $\eta$ by deploying the highest possible magnetic field in the largest available volume is actually common to the four families of magnetohydrodynamic thruster \cite{doragh1963magnetohydrodynamic,phillips1962prospects,Way1968,doss1990need,doss1991flow,doss1992overview}. However, equations (\ref{eq:effici0})-(\ref{eq:effici00}) also give $\mathcal{P}_e$  as \cite{Lin1991}
\begin{eqnarray}
\mathcal{P}_e=\left [1 - k(1+K_0)\, \eta \right] \frac{U_0^2}{R},
\end{eqnarray}
showing that $\mathcal{P}_e$ approaches $0$ when the second term of Eq~(\ref{eq:efficisi}) is minimized. Thus, maximizing $\eta$ gives a vanishing thrust. As pointed out in \cite{Lin1991}, rather than $\eta$, one should thus optimize $\mathcal{P}_m=\eta P_e$. Neglecting $K_0$, the value of $\eta$ which maximizes $\mathcal{P}_m$ is $\eta=1/(2k)$, which corresponds to a load factor $K=2$. With $k=1$, one thus should expect that $50 \%$ of the electrical power is consumed in Joule heating. 

Considering typical speed and size of commonly used ships, it is interesting to estimate how an MHD thruster can compete with usual propulsion method. Since its efficiency increases with $B$ (equation \eqref{eq:etaS}), one can thus determine the typical magnetic field required to obtain an acceptable efficiency for these ships. Using the typical ship length $L$ and velocity $u_d$, Eq~(\ref{eq:etaS}) gives
\begin{eqnarray}
B=\sqrt{\frac{\eta}{1-\eta} \frac{\rho}{2 \sigma \tau}} \label{eq:etaBB}
\end{eqnarray}
with $\mathcal{K} \sim \rho L^2/2$, $\mathcal{V} \sim L^3$, and where $\tau=L/u_d$ is the typical time corresponding to the time required for the ship to move for a distance equal to its length. Note that the field required by Eq~(\ref{eq:etaBB}) depends on the typical time scale only, and not on the ship length. Considering a ship of length $L = 10\, \textrm{m}$ traveling in seawater where $\rho= 10^3\, \textrm{kg.m}^{-3}$ and $\sigma = 5\, \textrm{S.m}^{-1}$, at a velocity $u_d =10\, \textrm{m.s}^{-1}$, leading to a typical time $\tau=1\, \textrm{s}$, Eq~(\ref{eq:etaBB}) gives a magnetic field $B=10\, \textrm{T}$ to maximize $\mathcal{P}_m$. The same magnetic field will be required for a small scale ship of size $10\, \textrm{cm}$ traveling at $u_d =10\, \textrm{cm.s}^{-1}$ since the typical time  $\tau=1\, \textrm{s}$ remains the same. Practical magnetic field intensity which can be achieved are typically smaller than $B=10\, \textrm{T}$, therefore the current MHD thrusters cannot compete with usual propulsion method. Conversely, maximum magnet field strength being typically $0.1-1\, \textrm{T}$, using the same typical values in Eq~(\ref{eq:etaS}) shows that one can expect an efficiency $\eta \approx 0.01-1 \%$.

\subsection{Analytical solutions for the thruster} \label{sec:analyTH}

Equations (\ref{eq:new})-(\ref{eq:meca22}) can be solved numerically but analytical solutions are much harder to obtain. In order to make analytical progress, the simplified electrical Eq~(\ref{eq:14_simplified}) is considered, the Darcy friction factor is supposed to be $f_D=64/Re$, which corresponds to a laminar flow in the thruster ($Re<2300$), and $C_d$ is assumed to be constant. This latter hypothesis allows to solve Eq~(\ref{eq:new}), leading to the solution (\ref{eq:u_inf}) for $\lambda$. Note also that this latter hypothesis is not needed to obtain analytical solution for the static thruster since Eq~(\ref{eq:new}) is irrelevant in this particular case. As shown in S3 Appendix, analytical solutions of the system of equations (\ref{eq:14_simplified})-(\ref{eq:meca22}) can then be obtained under these assumptions. 

Based on order of magnitude arguments, results on the different solutions can be obtained. For instance, $K$ being dimensionless implies that a typical field $B_{typ}=(U_0-E_0)/(k u_d H)$ exists for the thruster. Assuming a balance between $U_0-E_0$ and the  other terms of Eq~(\ref{eq:14_simplified}) gives $I \sim (U_0-E_0)/R$ and then $U_0-E_0 \sim k u_d^3 \mathcal{K}R/(U_0-E_0)$, allowing the evaluation of the typical flow velocity in the thruster $u_d$. Since $K>1$ gives a thruster and $K<1$ an electricity generator, it is thus expected that the solutions change for $K \sim 1$, i.e. when $B \sim B_{typ}$, with $B_{typ}= [(U_0-E_0)\mathcal{K}R/k^2]^{1/3}/H $.

For instance, when $B \ll B_{typ}$, the solutions can be reduced to (for $f_D=0$)
\begin{eqnarray}
    I_{B \rightarrow 0} & =& \frac{U_0-E_0}{R} \label{eq:I_simple} \\
  u_{d, {B \rightarrow 0}}  &=&   \sqrt{\frac{(U_0-E_0)BH}{\mathcal{K}R}} , \label{eq:ud_simple}
\end{eqnarray}
showing that $u_d$ increases with $B$. This limit corresponds to the limit  where the voltage $k u_d BH$ induced by the flow is negligible compared to $U_0-E_0$. The analytical solutions also show that $u_d$ decreases with $B$ for large $B \gg B_{typ}$, which shows the existence of an optimal field $B_{opt}$ for $u_d$ (which is thus bounded when $B$ is varied). This optimum is actually obtained when $K=2$, with the optimal field $B_{opt}=2^{-1/3}\, B_{typ} \approx B_{typ}$, which could be expected. This optimum is the same as the one obtained in section \ref{sec:finEQ} for $\mathcal{P}_m$, and the efficiency $\eta=1/(2k)$ is recovered for $E_0=0$ (see equation (17) in S3 Appendix).

As shown in S3 Appendix, the analytical solutions also predict that the  efficiency is maximized for a certain voltage $U_0$. Looking for $B$ and $U_0$ which simultaneously maximizes $u_d$ and $\eta$, respectively, gives $U_0=3 E_0$ and $B=[E_0\mathcal{K}R/k^2]^{1/3}/H$, with an efficiency of $\eta=1/(3k)$. 

\subsection{Measurements of the ship velocity} \label{sec:expeSpeed}

\begin{figure}[h!]
\centering
\includegraphics[width=10cm]{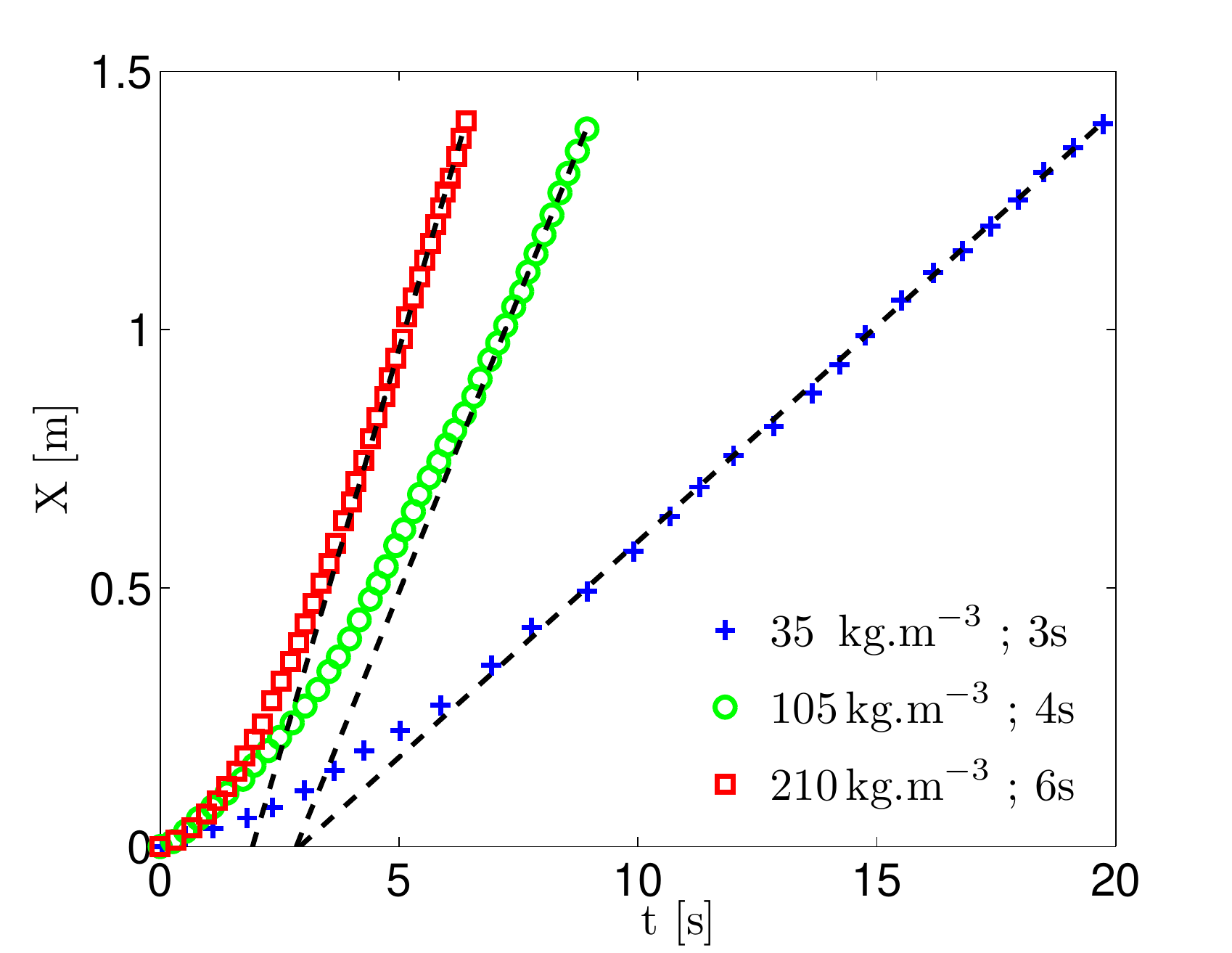} 
  \caption{{\bf Time evolution of the ship displacement for three different experimental configurations, varying salt concentration and battery power (symbols)}. Black-dashed lines represent linear fits over the last 6 points of each experiment. The terminal velocity of the boat is defined as the slope of the black-dashed line.}
         \label{fig08}
\end{figure}

Time evolution of the ship velocity $u_{\infty}$ is derived from video recording its displacement for the three batteries and six different salt concentrations, varying from the average seawater salinity ($\sim 35\, \textrm{g}.\textrm{L}^{-1}$) to the Dead Sea salinity ($\sim 300\, \textrm{g}.\textrm{L}^{-1}$). Fig~\ref{fig08} shows the displacement of the boat as a function of time for three different experiments. Increasing the salt concentration and battery voltage increases the ship velocity, but, for all configurations, a terminal constant velocity is obtained before reaching the end of the tank. 

In Table~\ref{tab2}, the measured current and terminal ship velocity are given for all the eighteen experimental runs. Note that the measured current is systematically small compared to the maximum current $I_m$ sustainable by the battery (given in Table~\ref{tab1}), insuring that the current is not limited by the battery but by the electrical circuit resistance. Table~\ref{tab2} also shows that, for a given battery, $u_{\infty}$ increases with the concentration of NaCl, but seems then to decrease for concentration $C > 210\, \textrm{kg.m}^{-3}$ (the presence of this maximum is expected, see section \ref{expe_elec}). Note also that the involved battery power is quite large, up to $1$ kW, for a maximum terminal velocity of $30\, \textrm{cm.s}^{-1}$. One can wonder if an internal inductive thruster would have given better performances. Considering an inductive internal thruster of typical size $10\, \textrm{cm}$, powered by a $200\, \textrm{A}$ superconducting magnet rotating at $50\, \textrm{Hz}$ and immersed in a similar fluid ($\rho=1100\, \textrm{kg.m}^{-3}$, $\sigma=10\, \textrm{S.m}^{-1}$), the typical velocity (resp. magnetic field) is expected to be $\approx 10\, \textrm{cm.s}^{-1}$ (resp. $\sim 0.1-1$ T), i.e. comparable to our slowest measurements \cite{hanfundamental2002}.

\begin{table}[!ht]
\begin{adjustwidth}{-2.25in}{0in} 
\centering
\caption{
{\bf Experimental measurements }}
\begin{tabular}{lll|llllll}
LiPo$^a$ & $U_0$ (V)  & $C$ (kg.m$^{-3}$) &  $I$ (A) & $\mathcal{P}_e$ (W) & $u_{\infty}$  (cm.s$^{-1}$) & \\
\hline
3s & $12.6$ & $35$ & $7.37$ & $92.86$ & $8.2$ \\
4s &$16.7$ & $35$ & $10.4$ & $173.7$ & $10.4$  \\
6s &$24.8$ & $35$ & $15.4$ & $381.9$ & $13.0$  \\
\hline
3s &$12.5$ & $70$ & $13$ & $162.5$ & $13.6$ \\
4s &$16.5$ & $70$ & $18.6$ & $306.9$ & $16.7$ \\
6s &$25$ & $70$ & $29$ & $725$ & $18.1$ \\
\hline
3s &$12.5$ & $105$ & $17.3$ & $216.2$ & $16.1$  \\
4s &$16.7$ & $105$ & $24.2$ & $404.1$ & $20.4$  \\
6s &$25$ & $105$ & $36.9$ & $922.5$ & $24.6$  \\
\hline
3s &$12.6$ & $175$ & $24$ & $302.4$ & $21.8$ \\
4s &$16.7$ & $175$ & $32$ & $534.4$ & $24.3$ \\
6s &$25.1$ & $175$ & $48$ & $1205$ & $24.6$ \\
\hline
3s &$12.6$ & $210$ & $23.9$ & $301.1$ & $23.7$ \\
4s &$16.7$ & $210$ & $33.4$ & $557.8$ & $22.6$ \\
6s &$25.1$ & $210$ & $49$ & $1230$ & $30.5$ \\
\hline
3s &$12.6$ & $291$ & $27$ & $340.2$ & $20.5$ \\
4s &$16.7$ & $291$ & $35.9$ & $599.5$ & $24.0$ \\
6s &$25.1$ & $291$ & $49$ & $1230$ & $23.3$ \\
\end{tabular}
\label{tab2}
\begin{flushleft} 
$^a$ Here, the control parameters are the kind of LiPo battery, which imposes the voltage $U_0$, and the salt concentration $C$.
\end{flushleft}
\end{adjustwidth}
\end{table}

Focusing on the terminal velocity, it is useful to collapse the results on a single curve making the comparison between the different configurations easier. To do so, it should be first noticed that, modifying the battery obviously changes the electric current $I$, but also the total mass $m=m_{ship}+m_{LiPo}$, thus the immersion depth $\mathcal{L}_y$ of the thruster. Hence, the ship velocity $u_{\infty}$ has to be expressed as a function of $I$, $m$ and $C$. 
The regular head loss being negligible in our experimental setup, Eq~(\ref{eq:meca22}) gives $u_d^2=IBH/\mathcal{K}$, leading to
\begin{eqnarray}
u_{\infty} = \lambda \, u_d=\lambda \sqrt{\frac{2\,  \overline{\sin \theta_2} \, IBH}{\rho S_{d} [\alpha_d -   \lambda^2 \alpha_{\infty}  + \xi(1-\lambda)^2]}}. \label{eq:coll}
\end{eqnarray}
For all different ship cross sections $S_w$ considered here, $2 \beta_d S_w C_d /S_d \approx 30 $ for our ship configuration, which is large compared to $\beta_{\infty} ^2$. Then, assuming a constant $C_d$, Eq~(\ref{eq:u_inf}) gives $\lambda \approx [2 \beta_d S_d/(S_w C_d)]^{1/2}$, which gives a typical value of $\lambda \approx 0.35$. This allows to simplify  Eq~(\ref{eq:coll}) into
\begin{eqnarray}
u_{\infty} \approx \sqrt{\frac{4\, IBH}{3 \rho S_w }} \approx \sqrt{\frac{4\, BH}{3\, \gamma } \frac{I}{m}}, \label{eq:shipp}
\end{eqnarray}
considering the limit $\lambda \ll 1$, and $\overline{\sin \theta_2}=1$, $\beta_d=\alpha_d=1$, $\xi \approx 2$, $C_d \approx 1$ (see Table~\ref{tab1}). The thruster solid cross section has been neglected in equation \eqref{eq:shipp} compared to the one of the ship, leading to $S_w \approx \gamma m/\rho$, where the constant $\gamma=(2 \mathcal{L}_z+D_y)/(2  \mathcal{L}_x^*  \mathcal{L}_y)$  only depends on the hull geometry (see also Eq~\ref{eq:massv}).

Eq~(\ref{eq:shipp}) shows that, at first order, the ship velocity does not depend on the fluid density $\rho$ or the thruster cross section $S_d$. It also shows that, at first order, the variable $I/m$ should allow a collapse of the data.

\begin{figure}[h!]
\centering
\includegraphics[width=10cm]{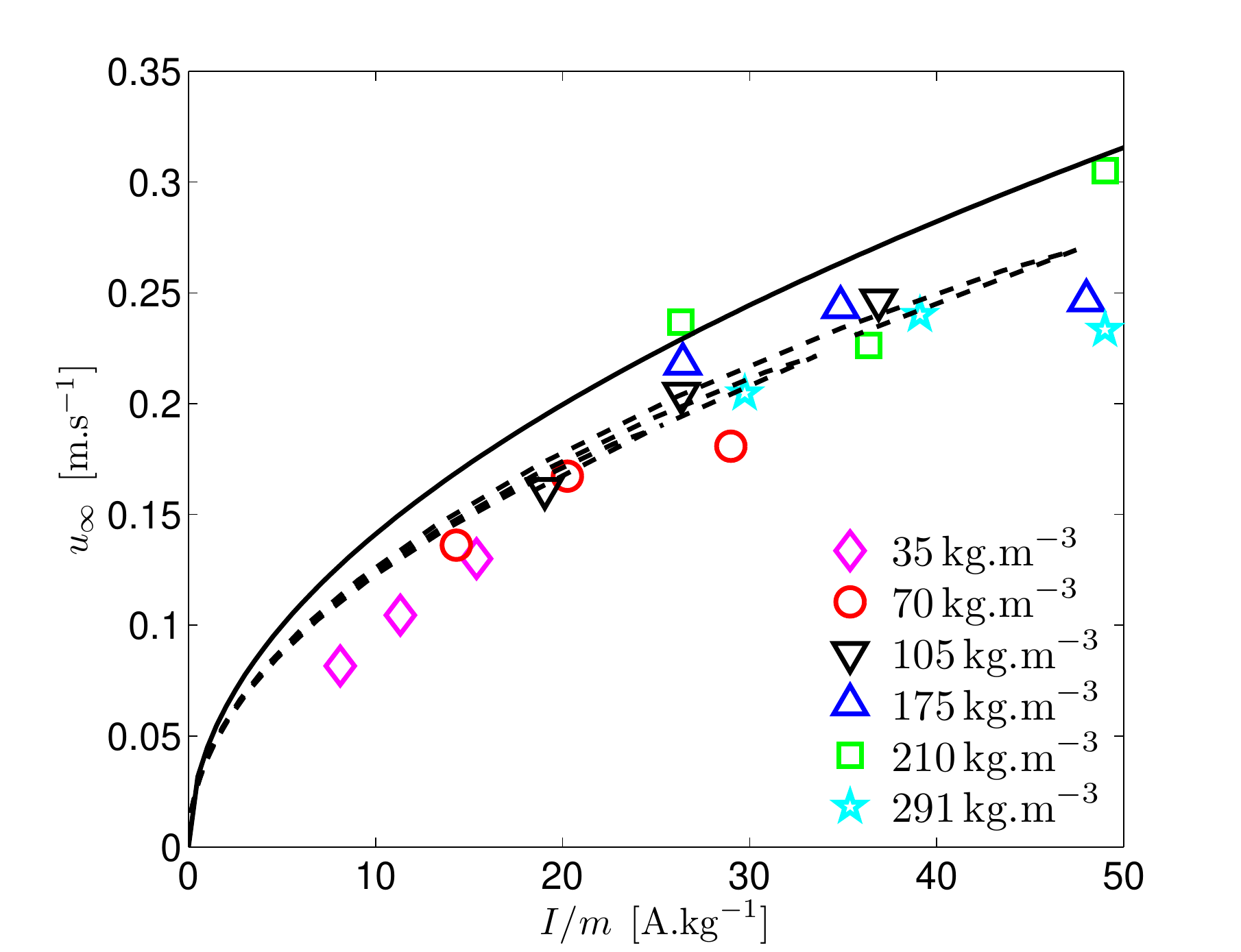} 
  \caption{{\bf Evolution of the ship velocity as a function of $I/m$. Experimental results for the three batteries (for each concentration) are represented by symbols.} The three dotted lines, which collapse rather well, correspond to equations (\ref{eq:new})-(\ref{eq:meca22}) for the three batteries. The solid line is given by Eq~(\ref{eq:shipp}).}
         \label{fig09}
\end{figure}

Experimental results for each concentration, for the three batteries, are presented in Fig~\ref{fig09} using the variable $I/m$. The terminal velocity has been obtained from the six last positions recorded by the camera (Fig~\ref{fig08}). The approximated Eq~(\ref{eq:shipp}) is plotted (solid line), as well as the exact theory, plotted for the three batteries (dashed lines), using equations  (\ref{eq:new})-(\ref{eq:meca22}), the values of Table~\ref{tab1}, and
\begin{eqnarray}
\frac{S_w-S_{th} }{2 \mathcal{L}_z+D_y}=\mathcal{L}_y= \frac{1}{2 \mathcal{L}_x^* \mathcal{L}_z} \left[ \frac{m}{(\rho_0+C)}-S_{th}L_x \right],
\end{eqnarray}
with $m=m_{ship}+m_{LiPo}$ and $\mathcal{L}_y$ being the immersion depth of the top of the thruster. As expected, the theoretical velocities collapse quite well for the three batteries. One can also notice a certain inflection point on the theoretical curves, corresponding to a concentration beyond which the velocity decreases when $C$ is increased. Without any adjustable parameters, the theory predictions of the maximum velocity reached by the ship are in good agreement with the experiments.


\section{MHD thruster optimization} \label{sec:DESIGN}

As presented in the previous sections, the developed theories are able to predict the main characteristics of our model ship. Nevertheless, even if some parts of its design were predicted in advance, other parameters such as the magnet size or the dimension of the U shape magnetic bridge were imposed by the manufacturer design of the commercially available components. Hence, it is legitimate to wonder if a more efficient ship could have been done and what the main characteristics of the ideal thruster should have been. In this section, based on expected orders of magnitude, a study on the optimal thruster is presented.

\subsection{Expected orders of magnitude} \label{sec:typic}

To design the MHD ship, it is useful to study the typical orders of magnitude we can expect for a small scale experimental setup. In order to predict these orders of magnitude a typical length of $10\, \textrm{cm}$ is chosen for the ship and Lithium-Polymer (LiPo) batteries as power source. 
Neodymium magnets are considered to generate the magnetic field. These magnets are the strongest type of permanent magnets commercially available and a good measure of their strength is given by their grades, defined as their maximum energy product. These grades usually range between N$35$ and N$52$, and can be related to $B_r$ using the empirical equation
\begin{eqnarray}
B_r \approx -0.00025597\, X^2 + 0.036314\, X + 0.22158, \label{eq:grade}
\end{eqnarray}
where $B_r$ is given in T, and where $X$ is the grade number (e.g. $40$ for a magnet of grade N$40$). A priori, the highest the grade, the more powerful the ship is, however, a high grade magnet is more inclined to physically break. Moreover, across the whole range of grades, $B_r$ only varies between $1.17\, \textrm{T}$ and $1.48\, \textrm{T}$. A good compromise between magnet strength and magnet solidity is the grade N$40$, one of the most common grade and the one used in this study. According to equation \eqref{eq:grade}, the residual magnetic field density obtained with these magnets is $B_r \approx 1.26\, \textrm{T}$. Then, using the Evershed criterion (see section \ref{sec:mag}) a typical field of half this value can be expected ($f=2$), thus  $ B \approx 0.63\, \textrm{T}$.
\noindent Using seawater as working fluid, typical values for density, $\rho \approx 10^3\, \textrm{kg}. \textrm{m}^{-3}$, minimum voltage needed for electrolysis $E_0 \approx 1\, \textrm{V}$, conductivity $\sigma \approx 10\, \textrm{S.m}^{-1}$, drag coefficient $C_d \approx 1$ can be chosen to obtain the orders of magnitude.

\begin{figure}[h!]
\centering
\includegraphics[width=12cm]{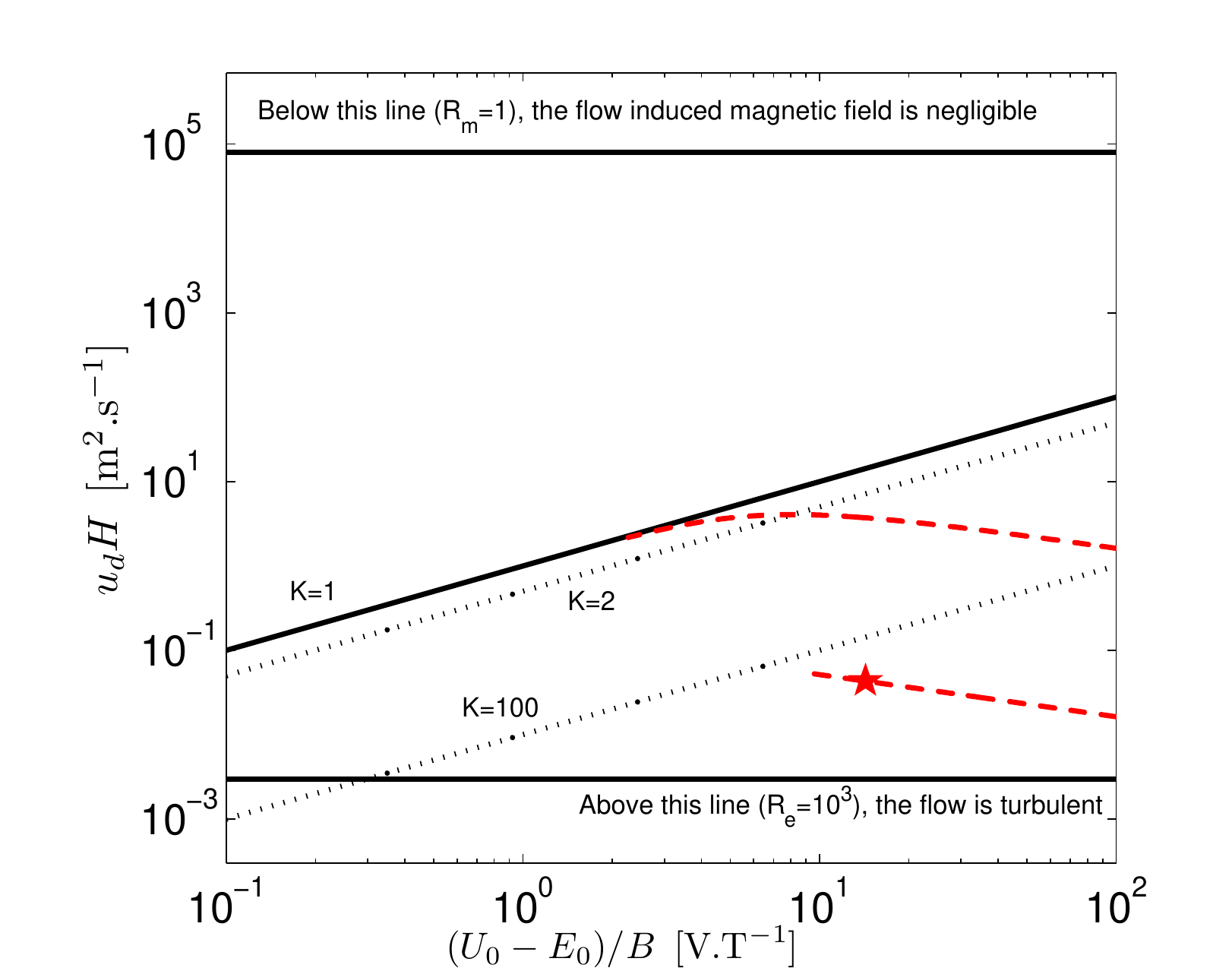} 
  \caption{{\bf Thruster working regimes (seawater, $k=1$)}. Below (resp. above) the thick solid tilted line $K=1$, the thruster behaves as a flow (resp. a voltage) generator. The dotted lines parallel to the thick line $K=1$ show solutions for constant values of $K$. Our typical thruster (the star) evolves along the thick dashed curve when $B$ is changed  in the range $B = 10^{-2} - 1\, \mathrm{T}$. The other thick dashed curve is similar, for a thruster size of $10$ m and a magnetic field from $B = 10^{-2}\, \mathrm{T}$ to $B= 4\, \mathrm{T}$.}
         \label{fig10}
\end{figure}

These typical values lead to a fluid electrical resistance $r=H/(\sigma l_x l_z) = 1\, \Omega$, which is large compared to the typical internal resistance $r_i \approx 0.01\, \Omega$ of a LiPo battery. For simplification, fringing effects are neglected so $R \approx r = 1\, \Omega$, and, assuming uniform flows in the thruster, $\lambda \approx 0.7$ which gives $\mathcal{K} \approx 3\, \textrm{kg.m}^{-1}$. Note that the thruster suction area is $S_{\infty}/S_d=1.4$ times larger than $S_d$, which constrains the experimental tank size. Using these parameters, equations (\ref{eq:I_simple}) and (\ref{eq:ud_simple}) give $I = 9\, \textrm{A}$ and $u_d \approx 43\, \textrm{cm.s}^{-1}$, i.e. $u_{\infty} \approx 30\, \textrm{cm.s}^{-1}$. This gives a load factor of $K \approx 330$ and a ship efficiency of $\eta = u_d BH/U_0 \approx 0.3 \%$. These expected orders of magnitude are actually confirmed by our experimental results (see section \ref{sec:expe}). The Reynolds number obtained using this values is $Re \approx 10^4$, which is in the range of validity for the Blasius estimate considered here (see section \ref{sec:thruster}). This also shows that the induced electric field ($ku_dBH \sim 0.03\, \textrm{V}$ with $k\sim 1$) is negligible ($RI\approx 9\, \textrm{V}$), and thus that the simplifying hypothesis $k=0$ is fully relevant here. We can therefor safely use equations  (\ref{eq:I_simple}) and (\ref{eq:ud_simple}).

In Fig~\ref{fig10}, axes have been chosen to gather a lot of information for an arbitrary thruster of size $H$, operating in a flow $u_d$. First, if follows from the definition of the load factor $K$ (Eq~\ref{eq:load0}) that the thick solid tilted line $K=1$ separates the two possible behaviors of the thruster, i.e. a flow or a voltage generator. Actually, the solutions for a given $K$ are represented by dotted lines parallel to the solid line $K=1$ (here, $K=2$ and $K=100$ are shown). The thick dashed lines represent how the solutions evolve when $B$ is varied, whereas the star corresponds to our typical MHD small scale ship model. It shows that increasing $B$ increases the velocity, as expected, until the maximum velocity reached for $K=2$ (see section \ref{sec:finEQ} or S3 Appendix). Beyond $K=2$, increasing $B$ decreases the velocity, even if the thruster efficiency continues to increase towards one (the solid curve tends along the line $K=1$). Fig~\ref{fig10} also shows that, in any cases, the induced magnetic field can be neglected for usual values of $(U_0-E_0)/B$. Note finally that our typical boat remains below the line $K=100$, even with very optimistic values of $B$, which confirms that the induced voltage can be neglected. However, this induced voltage cannot be neglected if we consider a ship of typical size $10\, \textrm{m}$ using a magnetic field of $B=4\, \textrm{T}$ (see solid curve of Fig~\ref{fig10}). Such a ship can actually represent the Yamato 1, a ship built in the early 1990s by Mitsubishi Heavy Industries \cite{motora1994development}. This prototype was able to reach speeds of $12\, \textrm{km.h}^{-1}$ with a thruster cross section size of $\sim 1\, \textrm{m}$ \cite{sasakawa1995superconducting,takezawa1995operation}. These values agree with the maximum $u_d H \approx 4\, \textrm{m}^2.\textrm{s}^{-1}$ shown in Fig~\ref{fig10}, which gives $u_d \approx 14\, \textrm{km.h}^{-1}$  for $H=1\, \textrm{m}$.

\subsection{Magnets distance maximizing the velocity} \label{sec:typic2}

Considering the experimental ship described in Table~\ref{tab1}, one can wonder whether, the distance between the magnets, $W$ has been well chosen. Varying $W$ will naturally change $B$, but also the ship mass and cross section $S_w$. Noting $m_{LiPo}$ the battery mass, $S_w$ can be calculated using the thruster cross section $S_{th}$ and the immersion depth $\mathcal{L}_y$ of the ship floats ($S_w=S_{th}+S_{hull}$, with the hull section area $S_{hull}$),
\begin{eqnarray}
\frac{S_w-S_{th} }{2 \mathcal{L}_z+D_y}=\mathcal{L}_y= \frac{1}{2 \mathcal{L}_x^* \mathcal{L}_z} \left[ \frac{m+\delta m}{(\rho_0+C)}-S_{th}L_x \right], \label{eq:massv}
\end{eqnarray}
where $m=m_{ship}+m_{LiPo}$ is the ship total mass, $S_{th}L_x$ is the volume of water displaced by the thruster only, and $\delta m=2.2\, (W-0.024)$ is the mass difference (in kg) for the ship when $W$ (in meters) varies. 

\begin{figure}[h!]
\centering
\includegraphics[width=10cm]{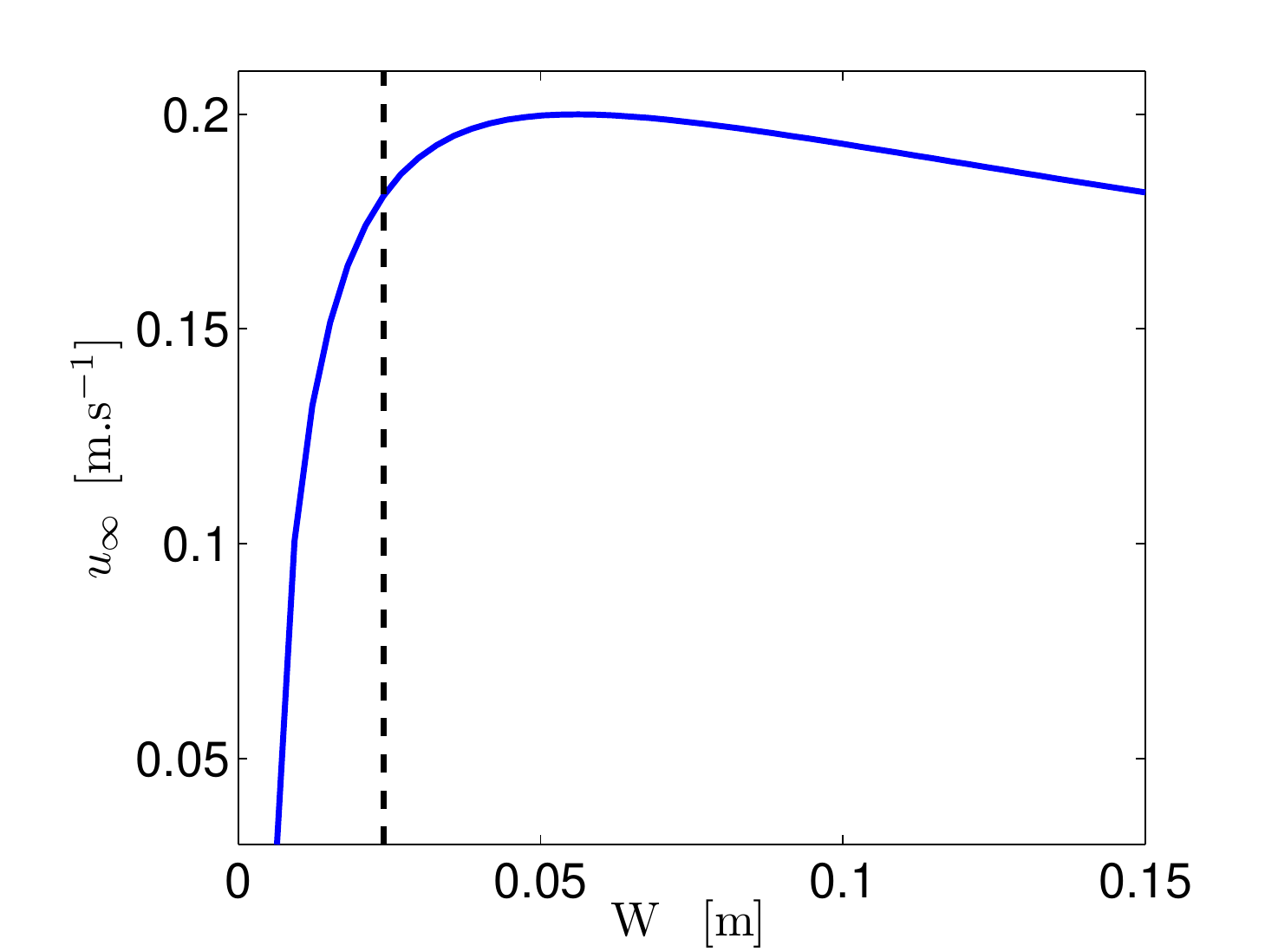}
  \caption{{\bf Ship velocity as a function of $W$.} For the intermediate battery 4s (the dashed line being the value for our experiments), with $\sigma \approx 8.7\, \textrm{S.m}^{-1}$, $B_r \approx 1.26\, \textrm{T}$ (skin friction modifications neglected). }
         \label{fig11}
\end{figure}

Using $l_z=W-\epsilon$, with $\epsilon=6\, \textrm{mm}$ the total thickness of the insulating material, equations (\ref{eq:elic22})-(\ref{eq:meca22}) can be solved using values in Table~\ref{tab1} and Eq~(\ref{eq:gain}) to take into account the $W$ dependency on the magnetic bridge effect, the mass of the ship, and $S_{th}$. Using $S_{th}=S_0+\epsilon_2 l_z$, where $S_0=960\, \textrm{mm}^2$ and $\epsilon_2=8\, \textrm{mm}$. The results, plotted in Fig~\ref{fig11}, show that an optimal $W$ exists.

To optimize the thruster on various parameters simultaneously, a simple estimate of this optimal $W$ is required. However, calculating $B$ as a function of $W$ requires to average in space the exact magnetic field  of a cuboidal magnet (given by \cite{camacho2013alternative}) which leads to complex calculations. A simpler expression of the magnetic field generated by the magnets can be obtained assuming that the magnetic field generated by a single magnet is given by the axial magnetic field of a cylindrical magnet of radius $a$ and length $L_z$. Then, the mean field $B$ between two cylindrical magnets can be estimated by (see S4 Appendix for details)
\begin{eqnarray}
B= B_r \frac{L_z}{a+W}. \label{eq:Bmeansol2}
\end{eqnarray}
Calculations with the exact magnetic field of a cuboidal magnet show that Eq~(\ref{eq:Bmeansol2}) correctly captures the evolution of $B$ with $W$ and hence is quite useful for rapid estimations. For instance, in \cite{Font2004}, $B$ is rather fitted by a quadratic polynomial in $W$, leading the authors to conclude erroneously that an optimum for $BW$ exists. Actually, the exact calculations of $B$ with the formulas of \cite{camacho2013alternative} confirm that $BW$ evolves as $B_r L_z W/(a+W)$, i.e. does not present any optimum.

Considering Eq~(\ref{eq:ud_simple}), with $R \approx H/(\sigma LW)$, and $\mathcal{K} \propto W$ (assuming $S_{d} \propto W$, and neglecting regular head losses), gives a constant $u_d$ for small $W$, but varies as $W^{-1/2}$  for large $W$. Equating the two asymptotic expressions shows that this change of behavior appears around $W \approx a$ for $r_i \approx 0$, which is the optimal $W$. Since $\lambda\sim W^{1/2}$ for small $W$ and $\lambda \sim 1$ for large $W$,  $u_{\infty}\sim W^{1/2}$ for small $W$, and $u_{\infty}$ varies as $W^{-1/2}$  for large $W$, exhibiting an optimum around $W_{opt} \approx a$.

\subsection{Interelectrode distance maximizing the velocity} \label{sec:typic3}

\begin{figure}[h!]
\centering
\includegraphics[width=10cm]{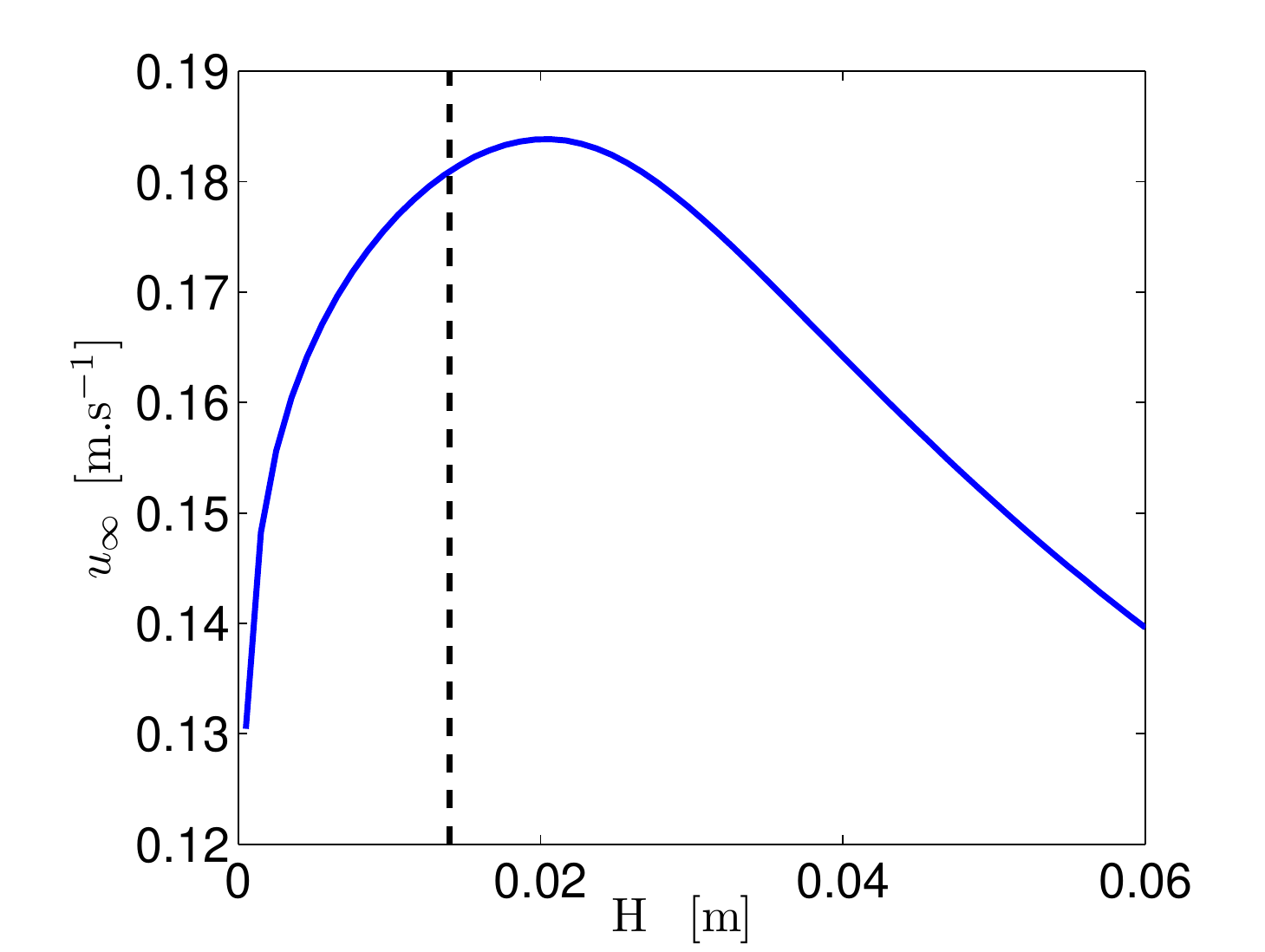} 
  \caption{{\bf Ship velocity as a function of $H$.} For the intermediate battery 4s (the dashed line being the design chosen for our experiments), with $\sigma \approx 8.7\, \textrm{S.m}^{-1}$, $B_r \approx 1.26\, \textrm{T}$  (skin friction modifications neglected). }
         \label{fig12}
\end{figure}

As for the distance separating the magnets, considering the values given in Table~\ref{tab1}, one can also wonder if the distance between electrodes $H$ has been well chosen. Varying $H$ will naturally change the current $I$, but also the thruster cross section $S_d$. The evolution of $u_{\infty}$ as a function of the separation distance between the electrodes $H$ can be calculated solving equations (\ref{eq:elic22})-(\ref{eq:meca22}) with the values in Table~\ref{tab1}. Using  $S_w=3268\, \textrm{mm}^2 $, corresponding to the case with 4s battery, the evolution of the ship velocity as a function of the distance between the electrodes can be calculated. Fig~\ref{fig12} shows this evolution and an optimum distance of $H \approx 2\, \textrm{cm}$ is obtained, maximizing $u_{\infty}$.

As for the magnetic field, to optimize the thruster on various parameters simultaneously, an estimate of this optimal $H$ is required. Considering the Eq~(\ref{eq:ud_simple}), with $R \approx H/(\sigma LW)$, and $\mathcal{K} \propto H$, assuming $S_{d} \propto H$, and neglecting regular head losses,  the flow velocity inside the thruster evolves as $u_d \sim H^{-1/2}$ for  small $H$. This unphysical behavior comes from the fact that the regular head losses have been neglected here but are actually not negligible at small $H$. Using the Huebscher law (\ref{eq:Hueb}) and the Hagen-Poiseuille law $f_D=64/Re$, valid for asymptotically small $H$, it can be shown that $u_d$ rather varies as $H^{1/4}$. If $H$ is larger than a critical value (around the magnet size $L_y$), the electric current flows in a zone where the magnetic field has an opposite direction, leading to an opposite Lorentz force direction. In this case, the mean field $B$ naturally decreases rapidly. Since $\lambda\sim W^{1/2}$ for small $W$ and $\lambda \sim 1$ for large $H$,  $u_{\infty}\sim W^{3/4}$ for small $H$, and decreases rapidly beyond typically $L_y$. To be sure to avoid the rapid decrease zone, we have chosen a value of $H$ slightly smaller than $L_y$, with  $H=1.4\, \textrm{cm} < L_y=2.4\, \textrm{cm}$, close to the optimum $H \approx 2\, \textrm{cm}$ represented in Fig~\ref{fig12}.

\subsection{Salt concentration maximizing velocities} \label{sec:typic4}

Since $C$ modifies $\sigma$ but also $\rho$, one can wonder if the concentration $C_{max}$, which maximizes the fluid conductivity $\sigma$ (see Eq~\ref{eq:Copt0}), is different from the optimal concentration maximizing the velocities $u_d$ or $u_{\infty}$. Noting $\rho_0$ the fluid density for $C=0$, the density reads as $\rho=\rho_0+C$, and thus $\mathcal{K}=\rho \Gamma=(\rho_0+C) \Gamma$, where $\Gamma$ is independent of $C$. Considering Eq~(\ref{eq:ud_simple}) in the particular case of vanishing internal resistance ($r_i=0$), $u_d$ is maximum for the concentration
\begin{eqnarray}
C_{opt}=  \frac{\left[\rho_0^{2/3} \left(a_0+\sqrt{\rho_0 b_0^2+a_0^2} \right)^{2/3} b_0^{1/3}-b_0 \rho_0 \right]^2}{\rho_0^{2/3} \left(a_0+\sqrt{\rho_0 b_0^2+a_0^2} \right)^{2/3} b_0^{4/3}}, \label{eq:Copt1}
\end{eqnarray}
which gives Eq~(\ref{eq:Copt0}) in the limit of large $\rho_0$. Using the values given in section \ref{sec:elec} for $a_0$, $b_0$, and $\rho_0=10^3\,  \textrm{kg.m}^{-3}$, Eq~(\ref{eq:Copt1}) gives  $C_{opt} \approx 176\,  \textrm{kg.m}^{-3}$, which is different from  $C_{max} \approx 197\,  \textrm{kg.m}^{-3}$. Solving numerically the full equations (\ref{eq:new})-(\ref{eq:meca22}) shows that $u_d$ is indeed maximum around $C \approx C_{opt}$. 

The fluid velocity $u_d$ inside the thruster does not dependent on the thruster immersion depth but the ship velocity $u_{\infty}$ is directly affected by the fluid drag force. The drag force being proportional to the ship cross-section $S_w$, which varies with $\rho$, and thus $C$, the optimum concentration for $u_d$ might not be the one maximizing the ship velocity $u_{\infty}$. In the parameters range considered in this work, this actually balances the $C$ dependency of  $\mathcal{K}=(\rho_0+C) \Gamma$, as shown by the estimate (\ref{eq:shipp}) of $u_{\infty}$ where the $C$ dependency only comes from $I$, i.e. from $\sigma$. One can thus finally expect that $u_{\infty}$ is maximum around  $C_{max} \approx 197\,  \textrm{kg.m}^{-3}$. Solving numerically equations (\ref{eq:new})-(\ref{eq:meca22}) gives a maximum $u_{\infty}$ for $C \approx 191\,  \textrm{kg.m}^{-3}$, in good agreement with our measurements (see Fig~\ref{fig09}). Note that, if the ship considered in this work was a submarine, $S_w$ would be constant, and the concentration maximizing  $u_{\infty}$  would rather be  $C_{opt} \approx 176\,  \textrm{kg.m}^{-3}$.

\section*{Conclusion}

Magnetohydrodynamics is a central part of Physics which governs many astrophysical or geophysical observations, such as flows in stellar layers, in planetary liquid cores or in accretion disks. On the other hand, very few MHD flows can be easily and directly observed in our daily life. 

This paper presents an experimental and theoretical study of a ship on salt water, self-propelled using magnetohydrodynamic forces. Despite the relative simplicity of the experimental setup, this is one of the first complete and self-consistent studies of a magnet/battery small scale MHD ship. The relevant theoretical equations were introduced for each component of the MHD thruster, followed by step-by-step experimental validations of the theory. This allowed us to validate theoretical predictions about the electrical properties of the fluid as a function of the voltage and the salt concentration (Tafel and Kohlrausch laws). Then, considering the hydrodynamics properties of both the thruster and the ship, and using the electrical and magnetic equations allowed the prediction of the ship velocity, without any adjustable parameters. Given the good agreement with the experimental results, the theory has then been used to optimize the different MHD ship parameters such as the distance separating the magnets or the electrodes. 

The experimental ship used in this work is actually well optimized for speed. With a typical power of $1000$ W and an efficiency of the order $0.1 \%$, our MHD ship is able to reach a maximum velocity of $0.3\,\textrm{m.s}^{-1}$. Note however that, given this poor efficiency, our ship only uses $1$ W for propulsion. For magnets based MHD ships, the strength of the magnets currently available actually limits the efficiency at this order of magnitude. To obtain the same efficiency than conventional propellers, MHD thrusters require compact and light generators of approximately $10\, \textrm{T}$ magnetic fields, which still remains challenging nowadays.

Despite the poor efficiency of such a propulsive method for commercial boats, small scale MHD ships are probably the easiest and the most recreational demonstration of the Lorentz force in a fluid. For this reason, such an experiment represents an ideal example case for undergrad students. 

\section*{Acknowledgments}
The authors gratefully acknowledge Yann Alexanian and Adeline Richard who participated to the preliminary studies. The authors also thank Julien Landel for discussions and suggestions that improved the paper writing.

\appendix   

\section{Discussion of electrolysis complications} \label{sec:electrocomp}
In many electrolysis reactions, a limiting current density, i.e. a maximum value for $j$ in Eq~(\ref{eq:Tafel}), is reached, indicating that the electrolysis have consummated all the reactant present in the thin diffusion layer in contact with the electrodes. Here, this limiting diffusion regime can only appear for $\textrm{Na}^+$ and $\textrm{Cl}^-$ because the other reactants are the solvent (water) and the electrodes \cite{mathon2009electro}. Moreover, for  $C=35\, \textrm{kg}.\textrm{m}^{-3}$, this limiting diffusion regime is reached for $j \sim 10^3 \, \textrm{A.m}^{-2}$ in a cell, whereas the limiting current density of a turbulent flowing electrolyte is rather $j \sim 10^7 \, \textrm{A.m}^{-2}$ , which is far above the typical current density $j \sim 10^3-10^4 \, \textrm{A.m}^{-2}$ considered in this work \cite{Boissonneau1999,Boissonneau1999a}. This maximum value for $j$ is thus not relevant for our MHD thruster.

Since the MHD thruster imposes a magnetic field in the fluid, one can wonder if the presence of a magnetic field modifies the picture. The Lorentz force acting on the charge carrier is actually modified by the presence of a magnetic field, giving an effective reduction in mobility for motion perpendicular to the magnetic field. The electrical conductivity in this direction (Pedersen conductivity) is thus reduced. This so-called magnetoresistance can be estimated by using a usual Drude model of electrical conduction, which shows that the resistance if increased by a factor $1+(\mu^* B)^2 $, where $\mu^*$ is the charge carrier mobility and $B$ the magnetic field (see \cite{tronel1982application} for details). In our experiments, $\mu^* B$ is typically of order $10^{-7}$ at most, and this effect is thus largely negligible.

Since the MHD thruster pumps water, one can also wonder how a non-zero fluid velocity in the electrolysis cell (i.e. a moving ship) modifies the picture, and how this affects the performances of the MHD thruster. First, the Tafel slope $A_0$ is increased by a non-zero fluid velocity, probably because of hydrodynamic boundary layer resistance enhancement \cite{petrick1992results}. Second, following \cite{picologlou1992experimental}, one can notice that the presence of bubbles downstream is dependent on both current density and fluid velocity. Indeed, hydrogen bubbles exist downstream only below a certain critical fluid velocity $u_c$, which increases with the imposed current density (e.g. $u_c=3.5\, \textrm{m}.\textrm{s}^{-1}$ for $j=250\, \textrm{A.m}^{-2}$, $u_c=7\, \textrm{m}.\textrm{s}^{-1}$ for $j=500\, \textrm{A.m}^{-2}$). It seems thus that the hydrogen goes into solution rapidly for large enough fluid velocities \cite{picologlou1992experimental}. Second, a large enough fluid velocity also allows to sweep the electrolysis bubbles downstream, avoiding to accumulate electrolysis bubbles on the electrodes. These bubbles can indeed form large insulating gas pockets, which would drastically reduce the electric current. The bubbles are $0.1-0.5\, \textrm{mm}$ in diameter, which gives a rise velocity in water of $1-5\, \textrm{cm}.\textrm{s}^{-1}$, and most of the bubbles are thus swept out of the channel before rising the top electrode as soon as the fluid velocity is larger than $\sim 0.2\, \textrm{m}.\textrm{s}^{-1}$ (see \cite{tempelmeyer1990electrical} for details). Note also that  \cite{Boissonneau1999} show that the electrolysis bubbles do not affect the flow. Since the fluid average velocity $u_d$ is systematically larger than $0.2\, \textrm{m}.\textrm{s}^{-1}$ in our measurements, electrolysis bubbles do not need to be taken into account.

\section{$\alpha$ and $\beta$ for the Hartmann flow} \label{app:hart}

The coefficients ($\alpha,\beta)$ for an arbitrary flow profile $u$ through a section $S$ are defined by  $\alpha=1/S \cdot \int_S [u/\bar{u}]^3 \textrm{d} \tau$ and  $\beta=1/S \cdot \int_S [u/\bar{u}]^2 \textrm{d} \tau$, with $\bar{u}$ the mean flow velocity). For instance, a Poiseuille flow in a cylinder gives $(\alpha=2,\beta=4/3)$, and for a plug flow, $\alpha \approx \beta \approx 1$.

\begin{figure}[h!]
\centering
\includegraphics[width=10cm]{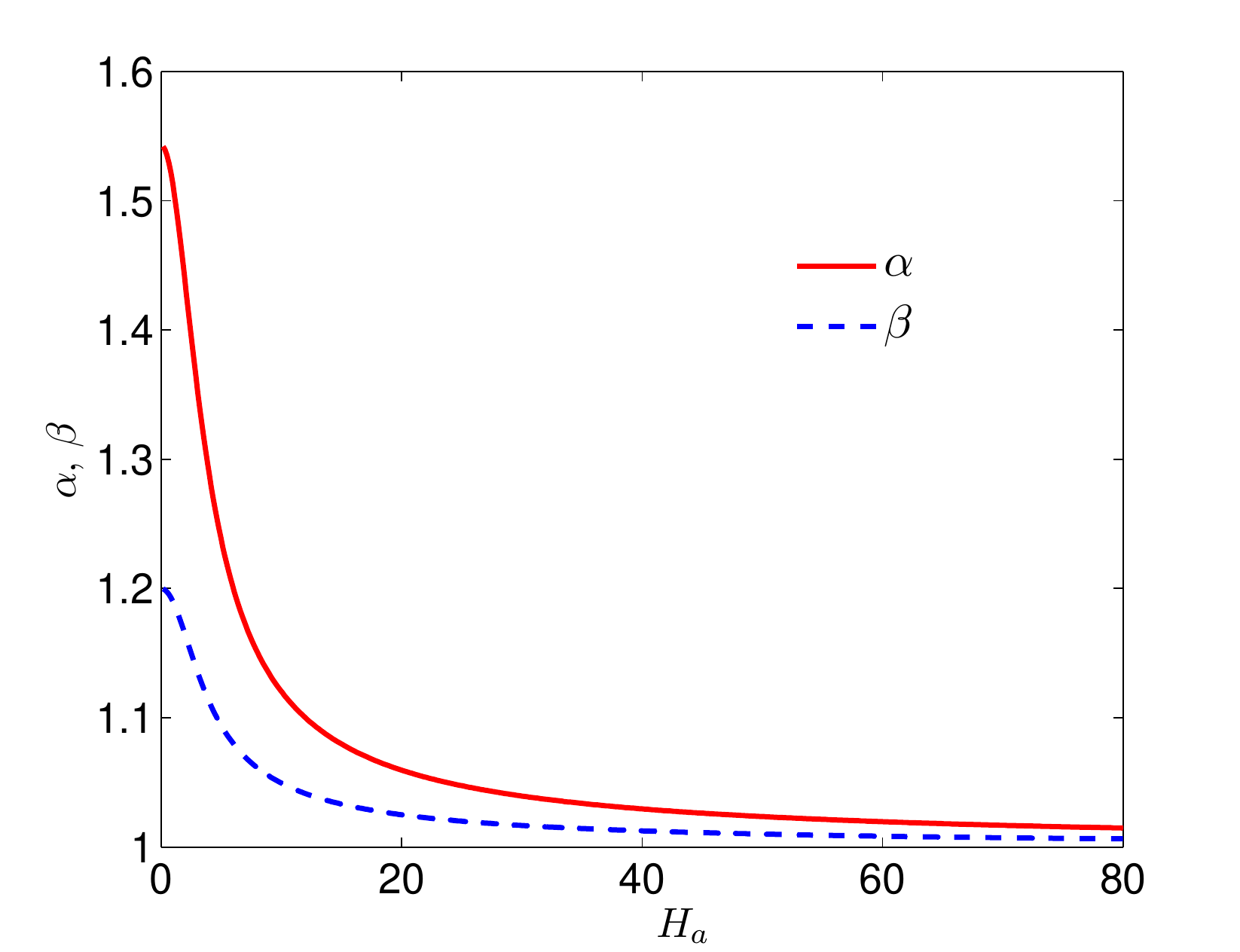} 
  \caption{{\bf Evolution of coefficients $\alpha$ and $\beta$ with $H_a$.}}
         \label{fig:coeff}
\end{figure}

To investigate the effect of a magnetic field on ($\alpha,\beta)$ in a simple manner, a plane Poiseuille flow is considered in presence of a magnetic field. This so-called Hartmann flow \cite{hartmann1937hg} is thus the flow between two parallel plates separated by a distance $W$, with a uniform magnetic field $B$ perpendicular to the planes. The velocity is then \cite{hartmann1937hg}
\begin{eqnarray}
\frac{u}{\bar{u}}=H_a\, \frac{\mathrm{cosh}(H_a)-\mathrm{cosh}(H_a \, Z)}{H_a\, \mathrm{cosh}(H_a)-\mathrm{sinh}(H_a)}, \label{eq:Hart}
\end{eqnarray}
where $Z=z/(W/2)$, using an axis $Oz$, perpendicular to the planes, with an origin located at a distance $W/2$ from the planes. The Hartmann number $H_a$ is given by $H_a=WB (\sigma/\eta)^{1/2}/2$, where $\sigma$ is the fluid electrical conductivity, and $\eta$ the fluid dynamic viscosity. Using the velocity (\ref{eq:Hart}) in  $\beta= 1/2 \cdot \int_{-1}^1 \left[u/\bar{u}\right]^2\ dZ$  gives
\begin{eqnarray}
\beta=\frac{H_a}{4}\frac{\sinh(2H_a)-8\cosh(H_a) \sinh(H_a)+Q}{H_a^2\cosh^2 H_a-H_a\sinh(2H_a)+\sinh^2 H_a}
\end{eqnarray}
with $Q=2H_a[1+2\cosh^2 H_a]$. Similarly, $\alpha=N H_a^2/(12D)$ is obtained from $\alpha= 1/2 \cdot \int_{-1}^1 \left[u/\bar{u}\right]^3\ dZ$, with $N=3H_a \cosh(3H_a)-\sinh(3H_a)+\cosh(H_a)[27H_a+9 \sinh(2H_a)]-\sinh(H_a) [27+18 \cosh(2H_a)]$ and 
$D=H_a \cosh(H_a) [H_a^2 \cosh^2 H_a+3 \sinh^2 H_a]-\sinh (H_a)[\sinh^2 H_a+3H_a^2 \cosh^2 H_a]$.

The evolution of both $\alpha$ and $\beta$ as a function of $H_a$ is represented in Fig~\ref{fig:coeff}. Typical values of $\alpha=54/35$ and $\beta=6/5$ for the plane Poiseuille flow are recovered for $H_a=0$. Both coefficients tend towards $1$ in the limit $H_a \gg 1$, where the flow tends to a uniform flow profile. In our case, a typical value is $H_a = 2.5$, obtained for $W=10\, \textrm{cm}$, $B=0.5\, \textrm{T}$, $\sigma=10\, \textrm{S.m}^{-1}$, and $\eta=10^{-3}\, \textrm{Pa.s}$.

\section{Analytical solutions for the thruster} \label{sec:analy}
We aim at solving analytically equations (\ref{eq:new})-(\ref{eq:meca22}) in order to obtain expressions for the thruster optimal parameters. To do so, we rely on the three assumptions described in section \ref{sec:analyTH}. Assuming a constant  $C_d$ allows to uncouple Eq~(\ref{eq:new}), leading to the solution (\ref{eq:u_inf}). Thus, $u_{\infty}$ is known as soon as $u_d$ is obtained. The MHD thruster is thus fully governed by equations (\ref{eq:elic22})-(\ref{eq:meca22}), and the two other assumptions allow to write them as
\begin{eqnarray}
    U_0 & =& E_0+RI+k u_d BH \label{eq1} \\
  IBH  &=&   \mathcal{K} \left(1+\tilde{\nu}/u_d \right)\, u_d^2 , \label{eq2}
\end{eqnarray}
where $\tilde{\nu}=64\, \nu\, L_x/[D_h^2\, (\alpha_d -   \lambda^2 \alpha_{\infty}  + \xi(1-\lambda^2))]$. Solving equations (\ref{eq1})-(\ref{eq2}) leads to
\begin{eqnarray}
   \frac{ I}{I_{B \rightarrow 0}} & =& 1- \left(2\kappa-\frac{\tilde{\nu}}{u_{d, {B \rightarrow 0}}} \right)(\sqrt{1+\kappa^2}-\kappa)  \\
  \frac{u_{d}}{u_{d, {B \rightarrow 0}}}  &=& \sqrt{1+\kappa^2}-\kappa, \label{eq:u_exact}
\end{eqnarray}
with 
 \begin{eqnarray}
\kappa=\frac{1}{2} \left( \frac{B}{B_{typ}} \right)^{3/2} + \frac{\tilde{\nu}}{2 \,  u_{d, {B \rightarrow 0}}},
\end{eqnarray}
where $B_{typ}= [(U_0-E_0)\mathcal{K}R/k^2]^{1/3}/H $, and
\begin{eqnarray}
    I_{B \rightarrow 0} & =& (U_0-E_0)/R \label{eq:IBl} \\
  u_{d, {B \rightarrow 0}}  &=&   \sqrt{(U_0-E_0)BH/(\mathcal{K}R)} . \label{eq:ud_simple2}
\end{eqnarray}
Equations (\ref{eq:IBl})-(\ref{eq:ud_simple2}) are solutions of equations (\ref{eq1})-(\ref{eq2}) when $\kappa=0$, i.e. when the regular head loss and induced electric field are negligible ($\tilde{\nu}=0$, $k=0$). In this limit, the equations are uncoupled: $I$ is fixed by Eq~(\ref{eq1}) and $u_d$ by Eq~(\ref{eq2}). In the other limit ($RI \ll k u_d BH$, $\tilde{\nu}=0$), the equations are also uncoupled and the thruster behaves as a current generator. Then $u_d$ is given by Eq~(\ref{eq1}), and $I$ by Eq~(\ref{eq2}), i.e.
\begin{eqnarray}
    u_{B \rightarrow \infty} & =& (U_0-E_0)/(kBH) \\
  I_{d, {B \rightarrow \infty}}  &=&  \mathcal{K} (U_0-E_0)^2/(k^2 (BH)^3) ,
\end{eqnarray}
which allows to give a physical interpretation to $\kappa$ with
\begin{eqnarray}
  2\, \kappa=\frac{  u_{d, {B \rightarrow 0}}}{ u_{B \rightarrow \infty} }=\sqrt{\frac{ I_{B \rightarrow 0}}{  I_{d, {B \rightarrow \infty}} }} .
\end{eqnarray}
When $B$ is increased,  $ u_{d, {B \rightarrow 0}}$ increases via the term $IBH$, and $ u_{B \rightarrow \infty}$ decreases because of the term $k u_d BH$. This shows that a magnetic field value $B_{opt}$ should maximise $u_d$. Solving $\partial_B u_d=0$ gives $B_{opt}$ as
 \begin{eqnarray}
B_{opt}=\left[ \frac{\mathcal{K}R(U_0-E_0)}{2 k^2 H^3}\right]^{1/3} =2^{-1/3}\, B_{typ}, \label{eq:Bopt0}
\end{eqnarray}
where we have assumed  $\tilde{\nu}=0$ (negligible regular head loss). Under this assumption, and for the field (\ref{eq:Bopt0})
 \begin{eqnarray}
  u_d(B_{opt}) = \max_B u_d = \left[ \frac{(U_0-E_0)^2}{4kR\mathcal{K}} \right]^{1/3}
\end{eqnarray}
which corresponds to $\kappa(B_{opt})=2^{-3/2} $, $K=2$ and
 \begin{eqnarray}
I(B_{opt}) = \frac{U_0-E_0}{2R}  \, \, \, ; \, \,  \, \eta(B_{opt})=\left[ \frac{1}{2k}-\frac{E_0}{2kU_0} \right]. \label{eq:optiIB}
\end{eqnarray}

One can also check if an optimum electric field exists, maximising the thruster efficiency. Solving $\partial_{U_0} \eta$, an explicit expression for $U_0^{\eta_{max}}$ is obtained. Assuming $\tilde{\nu}=0$, the expression reduces to
 \begin{eqnarray}
U_0^{\eta_{max}} = 2 E_0 + k \sqrt{E_0 (BH)^3/(\mathcal{K}R)},
\end{eqnarray}
leading to a maximum efficiency of 
 \begin{eqnarray}
\eta_{max} = \max_{U_0} \eta = \frac{1}{k+2\sqrt{E_0 \mathcal{K} R /(BH)^3}}.
\end{eqnarray}
For this particular voltage, the current and velocity are
 \begin{eqnarray}
I^{\eta_{max}} =E_0/R \, \, \, ; \, \,  \, \,  \, ,u_d^{\eta_{max}}=\sqrt{E_0 BH/(\mathcal{K}R)},
\end{eqnarray}
giving the load factor
 \begin{eqnarray}
K=1+\sqrt{E_0 \mathcal{K}R/(k^2 (BH)^3)}
\end{eqnarray}

Now, using $B_{opt}$, the couple $(B,U_0)$ which simultaneously maximise the velocity and the efficiency is
 \begin{eqnarray}
U_0^{max} = 3 E_0 \, \, \, \, \, ; \,  \,  \, B^{max}=(E_0 \mathcal{K}R)^{1/3}/(k^{2/3}H),
\end{eqnarray}
which corresponds to $\kappa(B_{opt})=2^{-3/2} $, $K=2$ and 
 \begin{eqnarray}
I^{\eta_{max}} =\frac{E_0}{R} \, \,  \, ; \, \,   \, u_d^{\eta_{max}}=\left(\frac{E_0^2}{k \mathcal{K}R} \right)^{1/3} \, \,  ; \, \,  \, \eta=\frac{1}{3k}
\end{eqnarray}

\section{Mean field of a cylindrical magnet} \label{sec:sol}

Focusing on the variation of magnetic field $B$ with the width of the thruster $W$, fringing effects can be neglected, and a simple model is chosen to study how $B$ evolves with $W$: the magnetic field is assumed to be the one on the main axis $Oz$ of a cylindrical magnet, i.e. of a solenoid, of radius $a$ and length $L_z$
\begin{eqnarray}
\frac{B_z}{B_r}= \frac{z/2+L_z/4}{\sqrt{a^2+(z+L_z/2)^2}} - \frac{z/2-L_z/4}{\sqrt{a^2+(z-L_z/2)^2}} , \label{eq:soleno}
\end{eqnarray}
where $B_r$ is the limit of (\ref{eq:soleno}) when $L_z  \rightarrow \infty$ (i.e. $\lim_{L_z \rightarrow \infty} B_z = B_r$). In this limit $L_z  \rightarrow \infty$, the field $B=B_r$ given by Eq~(\ref{eq:soleno}) corresponds to the uniform field within the magnet/solenoid  (for such an infinite magnet/solenoid, the field outside is simply $0$), which shows that $B_r$ is indeed the residual flux density (or induction) for a magnet (magnetic circuit closed at infinity, see section \ref{sec:mag}). 

Now, the magnet mean field $B=<B_z>$, which is $B_z$ averaged on a distance $W$, is 
\begin{eqnarray}
B &=& \frac{1}{W} \int_{L_z/2}^{W+L_z/2} B_z\,  \textrm{d}z \nonumber \\ &=& \frac{B_r}{2W} [a+\tilde{a}-\sqrt{a^2+L_z^2} - \sqrt{a^2+W^2} ], \label{eq:Bmeansol}
\end{eqnarray}
using $\tilde{a}=\sqrt{a^2+(W+L_z)^2)}$ and $B_z$ given by Eq~(\ref{eq:soleno}). Eq~(\ref{eq:Bmeansol}) shows that $B$ is constant for small $W$, but decreases as $1/W$ for large $W$. The critical value $W_c$ below which $B$ is nearly constant can be estimated by solving  $B_{(W \rightarrow 0)}=B_{(W \rightarrow \infty)}$. Noting $x=a/L_z$, we obtain $W_c/a=-x-1/x+\sqrt{1+x^2}+\sqrt{1+1/x^2}=f(x)$,  i.e. $W_c \approx a$  since $f(x)$ is nearly constant, around $1$ (which is confirmed by the limits for $x \rightarrow 0$ and $x \rightarrow \infty$, both equal to $1$).

In the usual limit $L_z \ll a$, a compact approximation of Eq~(\ref{eq:Bmeansol}) is 
\begin{eqnarray}
B= \frac{B_r}{2} \frac{L_z}{a+W},  \label{eq:Bmeansolapp}
\end{eqnarray}
which agrees with the two asymptotic expressions of Eq~(\ref{eq:Bmeansol}) for small and large $W$, and allows to recover $W_c \approx a$. Equations (\ref{eq:Bmeansol}), or its approximation (\ref{eq:Bmeansolapp}), are half the mean field created between two aligned attracting cylindrical magnets.


%
%
%



\begin{thebibliography}{}

\end{thebibliography}


\begin{thebibliography}{10}

\bibitem{weier2007flow}
Weier T, Shatrov V, Gerbeth G.
\newblock Flow control and propulsion in poor conductors.
\newblock In: Magnetohydrodynamics. Springer; 2007. p. 295--312.

\bibitem{rice1961propulsion}
Rice WA. Propulsion system; 1961.

\bibitem{friauf1961electromagnetic}
Friauf JB.
\newblock Electromagnetic ship propulsion.
\newblock Journal of the American Society for Naval Engineers.
  1961;73(1):139--142.

\bibitem{phillips1962prospects}
Phillips OM.
\newblock The prospects for magnetohydrodynamic ship propulsion.
\newblock Journal of Ship Research. 1962;43:43--51.

\bibitem{Way1968}
Way S.
\newblock Electromagnetic propulsion for cargo submarines. 1968;2(2):49--57.

\bibitem{swallom1991magnetohydrodynamic}
Swallom DW, Sadovnik I, Gibbs JS, Gurol H, Nguyen LV, Van~Den B, et~al.
\newblock Magnetohydrodynamic submarine propulsion systems.
\newblock Naval Engineers Journal. 1991;103(3):141--157.

\bibitem{Lin1995}
Lin TF, Gilbert JB.
\newblock Studies of helical magnetohydrodynamic seawater flow in fields up to
  twelve teslas.
\newblock Journal of Propulsion and Power. 1995;11(6):1349--1355.

\bibitem{nishigaki2000elementary}
Nishigaki K, Sha C, Takeda M, Peng Y, Zhou K, Yang A, et~al.
\newblock Elementary study on superconducting electromagnetic ships with
  helical insulation wall.
\newblock Cryogenics. 2000;40(6):353--359.

\bibitem{Khonichev1978}
Khonichev VI, Yakovlev VI.
\newblock Motion of a sphere in an infinite conductive fluid, produced by a
  variable magnetic dipole located within the sphere.
\newblock Journal of Applied Mechanics and Technical Physics.
  1978;19(6):760--765.

\bibitem{saji1988fundamental}
Saji Y, Iwata A, Sato M.
\newblock Fundamental studies of superconducting electromagnetic ship thruster
  driven by the alternating magnetic field.
\newblock Review of Kobe University of Mercantile Marine, Part II. 1988;36(7).

\bibitem{Khonichev1980a}
Khonichev VI, Yakovlev VI.
\newblock Motion of a plane plate of finite width in a viscous conductive
  liquid, produced by electromagnetic forces.
\newblock Journal of Applied Mechanics and Technical Physics.
  1980;21(1):77--84.

\bibitem{Shatrov1981}
Shatrov VI, Yakovlev VI.
\newblock Change of hydrodynamic drag of a sphere set in motion by
  electromagnetic forces.
\newblock Journal of Applied Mechanics and Technical Physics.
  1981;22(6):817--823.

\bibitem{yakovlev1980theory}
Yakovlev V.
\newblock Theory of an induction MHD propeller with a free field.
\newblock Journal of Applied Mechanics and Technical Physics.
  1980;21(3):376--384.

\bibitem{Convert1995}
Convert D.
\newblock Propulsion magnetohydrodynamique en eau de mer; 1995.

\bibitem{saji1978basic}
Saji Y, Kitano M, Iwata A.
\newblock Basic study of superconducting electromagnetic thrust device for
  propulsion in seawater.
\newblock In: Advances in Cryogenic Engineering. Springer; 1978. p. 159--169.

\bibitem{iwata1980construction}
Iwata A, Saji Y, Sato S.
\newblock Construction of model ship ST--500 with superconducting
  electromagnetic thrust system.
\newblock Proc ICEC. 1980;8:775--784.

\bibitem{motora1994development}
Motora S, Takezawa S.
\newblock Development of MHD ship propulsion and results of sea trials of an
  experimental ship YAMATAO--1.
\newblock In: Int. Conf. on Energy Transfer in MHD Flows, Aussois, France;
  1994. p. 501--510.

\bibitem{sasakawa1995superconducting}
Sasakawa Y, Takezawa S, Sugawara Y, Kyotani Y.
\newblock The superconducting MHD-propelled ship YAMATO-1. 1995;.

\bibitem{Yan2000}
Yan L, Sha C, Zhou K, Peng Y, Yang A, Qin J.
\newblock Progress of the MHD ship propulsion project in China.
\newblock IEEE transactions on applied superconductivity. 2000;10(1):951--954.

\bibitem{yan2000development}
Yan L, Wang Z, Xue C, Gao Z, Zhao B.
\newblock Development of the superconducting magnet system for HEMS-1 MHD model
  ship.
\newblock IEEE transactions on applied superconductivity. 2000;10(1):955--958.

\bibitem{Khonichev1981}
Khonichev VI, Iakovlev VI.
\newblock Theory of a free-field conduction propulsion unit.
\newblock {PMTF} Zhurnal Prikladnoi Mekhaniki i Tekhnicheskoi Fiziki.
  1981;21:109--118.

\bibitem{shatrov1985hydrodynamic}
Shatrov V, Yakovlev V.
\newblock Hydrodynamic drag of a ball containing a conduction-type source of
  electromagnetic fields.
\newblock Journal of Applied Mechanics and Technical Physics.
  1985;26(1):19--24.

\bibitem{pohjavirta1991feasibility}
Pohjavirta A, Kettunen L.
\newblock Feasibility study of an electromagnetic thruster for ship propulsion.
\newblock IEEE transactions on magnetics. 1991;27(4):3735--3742.

\bibitem{convert1995external}
Convert D, Thibault J.
\newblock External MHD propulsion.
\newblock Magnetohydrodynamics. 1995;31:290--297.

\bibitem{Font2004}
Font GI, Dudley SC.
\newblock Magnetohydrodynamic propulsion for the classroom.
\newblock The Physics Teacher. 2004;42(7):410--415.

\bibitem{Clancy1984}
Clancy T.
\newblock The Hunt for Red Ocotber.
\newblock HarperCollins; New Ed edition (2 Feb. 1998); 1984.

\bibitem{Baumgartl1993}
Baumgartl J, Hubert A, M{\"u}ller G.
\newblock The use of magnetohydrodynamic effects to investigate fluid flow in
  electrically conducting melts.
\newblock Physics of Fluids A: Fluid Dynamics (1989-1993).
  1993;5(12):3280--3289.

\bibitem{Thess2006}
Thess A, Votyakov EV, Kolesnikov Y.
\newblock Lorentz force velocimetry.
\newblock Physical Review Letters. 2006;96(16):164501.

\bibitem{Priede2011}
Priede J, Buchenau D, Gerbeth G.
\newblock Contactless electromagnetic phase-shift flowmeter for liquid metals.
\newblock Measurement Science and Technology. 2011;22(5):055402.

\bibitem{Brown1990}
Brown SH, Walker JS, Sondergaard NA, Reilly PJ, Bagley DE. Propulsive
  Efficiencies of Magnetohydrodynamic Submerged Vehicular Propulsors; 1990.
\newblock Available from:
  \url{http://oai.dtic.mil/oai/oai?verb=getRecord&metadataPrefix=html&identifier=ADA221088}.

\bibitem{Gilbert1991a}
Gilbert JB, Lin TF. Studies of {MHD} propulsion for underwater vehicles and
  seawater conductivity enhancement; 1991.
\newblock Available from:
  \url{http://oai.dtic.mil/oai/oai?verb=getRecord&metadataPrefix=html&identifier=ADA231623}.

\bibitem{thibault1994status}
Thibault J.
\newblock Status of MHD ship propulsion.
\newblock In: 2nd International Conference on Energy Transfer in MHD Flows,
  Aussois; 1994.

\bibitem{tempelmeyer1990electrical}
Tempelmeyer KE.
\newblock Electrical Characteristics of a Seawater MHD Thruster.
\newblock DTIC Document; 1990.

\bibitem{imageJ2012}
Schneider, C.A. and Rasband, W.S. and Eliceiri, K.W.
\newblock NIH Image to ImageJ: 25 years of image analysis
\newblock Nature methods.
2012;9(7):671--675.

\bibitem{Mitchell1988}
Mitchell D, Gubser D.
\newblock Magnetohydrodynamic ship propulsion with superconducting magnets.
\newblock Journal of Superconductivity. 1988;1(4):349--364.

\bibitem{Boissonneau1999b}
Boissonneau P.
\newblock Magnetohydrodynamics propulsion: a global approach of an inner DC
  thruster.
\newblock Energy conversion and management. 1999;40(17):1783--1802.

\bibitem{wright2007introduction}
Wright MR.
\newblock An introduction to aqueous electrolyte solutions.
\newblock John Wiley \& Sons; 2007.

\bibitem{JR9370000574}
Robinson RA, Davies CW.
\newblock 128. The conductivity of univalent electrolytes in water.
\newblock J Chem Soc. 1937; p. 574--577.
\newblock doi:{10.1039/JR9370000574}.

\bibitem{leus2004fringing}
Leus V, Elata D.
\newblock Fringing field effect in electrostatic actuators.
\newblock Technion-Israel Institute of Technology Technical Report No
  ETR-2004-2. 2004;.
  

\bibitem{bennett1978site}
Bennett JE.
\newblock On-site generation of hypochlorite solutions by electrolysis of
  seawater.
\newblock In: Available from Copyright Clearance Center, Inc., New York. In:
  Water--1977, AIChE Symposium Series. vol.~74; 1978.

\bibitem{leroy1978time}
Leroy R, Janjua M, Renaud R, Leuenberger U.
\newblock Time variations effects in unipolar water electrolyzers and their
  implications for efficiency improvement.
\newblock In: Proceedings of the Symposium on Water Electrolysis. vol.~78;
  1978. p.~4.
  
\bibitem{Boissonneau1999}
Boissonneau P, Thibault J.
\newblock Experimental analysis of couplings between electrolysis and
  hydrodynamics in the context of MHD in seawater.
\newblock Journal of Physics D: Applied Physics. 1999;32(18):2387.

\bibitem{atkins2006atkins}
Atkins P, De~Paula J.
\newblock Physical chemistry.
\newblock New York. 2006; p.~77.

\bibitem{petrick1992results}
Petrick M, Libera J, Bouillard J, Pierson E, Hill D.
\newblock Results from a large-scale MHD propulsion experiment.
\newblock Argonne National Lab.; 1992.

\bibitem{camacho2013alternative}
Camacho J, Sosa V.
\newblock Alternative method to calculate the magnetic field of permanent
  magnets with azimuthal symmetry.
\newblock Revista mexicana de f{\'\i}sica E. 2013;59(1):8--17.

\bibitem{yang1990potential}
Yang Z, Johansen T, Bratsberg H, Helgesen G, Skjeltorp A.
\newblock Potential and force between a magnet and a bulk Y1Ba2Cu3O7-$\delta$
  superconductor studied by a mechanical pendulum.
\newblock Superconductor Science and Technology. 1990;3(12):591.

\bibitem{engel2005calculation}
Engel-Herbert R, Hesjedal T.
\newblock Calculation of the magnetic stray field of a uniaxial magnetic
  domain.
\newblock Journal of Applied Physics. 2005;97(7):074504.

\bibitem{Schlichting}
Schlichting H.
\newblock Boundary-layer theory.
\newblock New York [etc.: McGraw-Hill; 1968.

\bibitem{Hoerner1965a}
Hoerner SF.
\newblock Fluid-dynamic drag.
\newblock 2nd ed. Midland Park, NJ: Verf.; 1965.

\bibitem{Tuck2002}
Tuck EO, Scullen DC, Lazauskas L.
\newblock Wave patterns and minimum wave resistance for high-speed vessels.
\newblock In: 24th Symposium on Naval Hydrodynamics, Fukuoka, Japan. vol. 813.
  Citeseer; 2002.Available from:
  \url{http://citeseerx.ist.psu.edu/viewdoc/download?doi=10.1.1.495.6935&rep=rep1&type=pdf}.
  
\bibitem{tuck1966shallow}
Tuck E.
\newblock Shallow-water flows past slender bodies.
\newblock Journal of fluid mechanics. 1966;26(01):81--95.

\bibitem{Davidson1992}
Davidson PA.
\newblock The role of pressure forces in {MHD} propulsion of submersibles.
  1992;104(5):39--42.

\bibitem{Huebscher1948}
Huebscher R.
\newblock Friction equivalents for round, square and rectangular ducts.
\newblock ASHVE Transactions (renamed ASHRAE Transactions). 1948;54:101--144.

\bibitem{muller2002liquid}
M{\"u}ller, U. and B{\"u}hler, L.
\newblock Liquid Metal Magneto-Hydraulics Flows in Ducts and Cavities.
\newblock Magnetohydrodynamics. 2002;1--67.

   \bibitem{van2005drag}
 Van den Berg, Thomas H and Luther, Stefan and Lathrop, Daniel P and Lohse, Detlef
\newblock Drag reduction in bubbly Taylor-Couette \newblock Physical review letters.
\newblock 2005; 94(4):044501.

\bibitem{doragh1963magnetohydrodynamic}
Doragh R.
\newblock Magnetohydrodynamic ship propulsion using superconducting magnets.
\newblock SNAME Trans. 1963;71:370--386.

\bibitem{doss1990need}
Doss E, Geyer H.
\newblock The need for superconducting magnets for MHD seawater propulsion.
\newblock Argonne National Lab., IL (USA); 1990.

\bibitem{doss1991flow}
Doss E, ROY G.
\newblock Flow characteristics inside MHD seawater thrusters.
\newblock Journal of Propulsion and Power. 1991;7(4):635--641.

\bibitem{doss1992overview}
Doss ED, Geyer HK.
\newblock An overview of MHD seawater thruster performance and loss mechanisms.
\newblock SAE Technical Paper; 1992.

\bibitem{Lin1991}
Lin TF, Gilbert JB, Roy GD.
\newblock Analyses of magnetohydrodynamic propulsion with seawater for
  underwater vehicles. 1991;7(6):1081--1083.

\bibitem{hanfundamental2002}
Han J, Sha C, Peng Y.
\newblock Fundamental Study on Alternating Magnetic Field MHD Thruster.
\newblock In: The 17th International Conference on Magnetically Levitated
  Systems and Linear Drives (MAGLEV'2002); 2002.

\bibitem{takezawa1995operation}
Takezawa S, Tamama H, Sugawawa K, Sakai H.
\newblock Operation of the Thruster for Superconducting
  Electromagnetohydrodynamic Propu1sion Ship" YAMATO 1".
\newblock Bulletin of the MESJ. 1995;23(1):46.

\bibitem{mathon2009electro}
Mathon P, Nouri A, Alemany A, Chopart J, Sobolik V, Baaziz D.
\newblock Electro-chemical processes controlled by high magnetic fields:
  application to MHD sea water propulsion.
\newblock Magnetohydrodynamics c/c of Magnitnaia Gidrodinamika.

\bibitem{Boissonneau1999a}
Boissonneau P, Thibault JP.
\newblock Sea water MHD: electrolysis and gas production in flow.
\newblock In: Transfer Phenomena in Magnetohydrodynamic and Electroconducting
  Flows. Springer; 1999. p. 251--266.

\bibitem{tronel1982application}
Tronel-Peyroz E, Olivier A.
\newblock Application of the Boltzman equation to the study of electrolytic
  solution in the presence of electric and magnetic fields.
\newblock Physico-Chemical Hydrodynamics. 1982;3:251--265.

\bibitem{picologlou1992experimental}
Picologlou B, Doss E, Black D, Sikes WC.
\newblock Experimental determination of magnetohydrodynamic seawater thruster
  performance in a two Tesla test facility.
\newblock SAE Technical Paper; 1992.

\bibitem{hartmann1937hg}
Hartmann J, Lazarus F.
\newblock Hg dynamics.
\newblock Levin \& Munksgaard; 1937.

\end{thebibliography}
\end{document}